\documentclass[11pt,a4paper]{article}
\usepackage{xy,amsmath,amssymb,slashed,url,bm,textgreek,upgreek}
\usepackage[table]{xcolor}
\usepackage{bbm,array,amsfonts,wrapfig,lscape,float,mathtools,multirow,longtable}
\usepackage{graphicx}
\usepackage{epstopdf}
\usepackage{braket}
\usepackage{tikz-cd}
\usepackage{newtxtext,newtxmath}
\def\be{\begin{equation}}
\def\ee{\end{equation}}

\def\hat{\widehat}

\def\tilde{\widetilde}

\def\[{\bigl [}

\def\]{\bigr ]}

\def\Z{{\mathbb Z}}

\def\tilde{\widetilde}
\def\sing{{\mathrm{sing}}}
\def\bar{\overline}

\font\teneurm=eurm10 \font\seveneurm=eurm7  \font\fiveeurm=eurm5
\newfam\eurmfam
\textfont\eurmfam=\teneurm \scriptfont\eurmfam=\seveneurm
\scriptscriptfont\eurmfam=\fiveeurm

\font\teneusm=eusm10 \font\seveneusm=eusm7 \font\fiveeusm=eusm5
\newfam\eusmfam
\textfont\eusmfam=\teneusm \scriptfont\eusmfam=\seveneusm
\scriptscriptfont\eusmfam=\fiveeusm

\font\tencmmib=cmmib10 \skewchar\tencmmib='177
\font\sevencmmib=cmmib7 \skewchar\sevencmmib='177
\font\fivecmmib=cmmib5 \skewchar\fivecmmib='177
\title{
\fontsize{18pt}{20pt}\selectfont\bfseries
Categorified Structures over Moduli Spaces:\\[0.5em]
\fontsize{14pt}{16pt}\selectfont\itshape
Anomalies, Non-Invertible Symmetries, and Exceptional Holonomy
}

\author[a]{Alonso Perez-Lona,}   
\author[a]{Eric Sharpe,}         
\author[a]{and Xingyang Yu}

\affiliation[a]{Department of Physics MC 0435, 850 West Campus Drive, Virginia Tech, Blacksburg, VA 24061}

\emailAdd{aperezl@vt.edu}
\emailAdd{ersharpe@vt.edu}
\emailAdd{xingyangy@vt.edu}

\abstract{In this note, we propose an extension of the relation between worldsheet global symmetries and structures over moduli spaces of superconformal field theories (SCFTs) to include noninvertible symmetries. The most familiar examples of such structures associated to an ordinary symmetry are the Bagger–Witten line bundles, which arise over the moduli spaces of two-dimensional $\mathcal{N} = (2,2)$ SCFTs from a non-anomalous worldsheet $U(1)_R$ symmetry and its associated spectral flow operators. Generalizing this setting, we consider examples involving anomalous worldsheet symmetries, which, despite not being gaugeable, can still give rise to global structures over moduli space—as illustrated by the momentum/winding symmetries in toroidal compactifications-and higher group gauge symmetry structure in spacetime. Motivated by this analogy, we conjecture the existence of a stack of fusion categories over moduli spaces of $G_2$ and Spin$(7)$ holonomy manifolds which acts as a noninvertible analogue of the Hodge or Bagger–Witten line bundles over Calabi–Yau moduli spaces. This proposal is based on the observation that SCFTs associated with such exceptional holonomy manifolds contain (tricritical) Ising sectors that play a role analogous to the $U(1)_R$ symmetry in $\mathcal{N} = 2$ theories. Although these symmetries are not gaugeable, they behave similarly to anomalous invertible symmetries, providing the conceptual foundation for the proposed moduli space structure.}
\begin{document}

\maketitle

\flushbottom

\section{Introduction}
\label{sec:intro}

Let $\mathcal{M}$ be a moduli space of conformal field theories (CFTs), each equipped with a global symmetry group $G$. One naturally expects the existence of a principal $G$-bundle $\mathcal{P}_G \to \mathcal{M}$, whose transition functions\footnote{By passing from $\mathcal{M}$ to a suitable stack over $\mathcal{M}$, one can arrange for triple overlap conditions to hold, so we suppress such constraints in this article.} are determined by global symmetry transformations relating CFTs on overlapping coordinate patches of $\mathcal{M}$.

Well-studied examples of such a structure include the Bagger–Witten line bundles, originally introduced in \cite{Witten:1982hu} to address a subtlety in moduli space geometry in $\mathcal{N}=1$ supergravity. These line bundles reappear in two-dimensional $\mathcal{N}=2$ SCFTs \cite{Periwal:1989mx,Distler:1992gi}, where their origin lies in the presence of a global $U(1)_R$ symmetry. They play a central role in both physics and mathematics: in four-dimensional supergravity, their first Chern class controls the K\"{a}hler form of the moduli space, the superpotential transforms as a section of the tensor square, and Fayet–Iliopoulos terms correspond to choices of equivariant structures \cite{Distler:2010zg,Hellerman:2010fv}. Mathematically, the tensor square of a Bagger–Witten line bundle is the Hodge line bundle of holomorphic top-forms (the relative canonical bundle) over Calabi–Yau moduli space. See \cite{Sharpe:2024rwb} for a recent overview of their role in SCFTs and supergravity.

More broadly, the Bagger–Witten bundle exemplifies how global symmetries on the worldsheet become local gauge symmetries in spacetime. In string compactifications, the worldsheet $U(1)_R$ symmetry induces a $U(1)$ gauge field in the low-energy theory, and the associated line bundle over moduli space reflects this emergent gauge structure.

\vspace{0.5em}

Given the central role of the $U(1)_R$ symmetry in generating such geometric structures, it is natural to ask whether analogous structures exist in more intricate compactifications—particularly those on manifolds of exceptional holonomy, such as $G_2$ and $\mathrm{Spin}(7)$. In these cases, there is no continuous $U(1)$ R-symmetry; instead, the only surviving invertible symmetry is a discrete $\mathbb{Z}_2$, corresponding to fermion parity. This $\mathbb{Z}_2$ symmetry, sometimes interpreted as the discrete $R$-symmetry of the two-dimensional $\mathcal{N}=1$ algebra, would at most give rise to a flat $\mathbb{Z}_2$ bundle over moduli space.

However, this $\mathbb{Z}_2$ is embedded within a larger non-invertible symmetry structure. SCFTs associated to $G_2$ and $\mathrm{Spin}(7)$ compactifications contain rational sectors governed by minimal models, as first noted in \cite{Shatashvili:1994zw}. In this paper, we identify those sectors as giving rise to non-invertible symmetries—described by
\begin{equation}
\begin{split}
	\mathrm{Spin}(7)&: \quad \text{Ising fusion category},\\
	G_2&: \quad \text{Tricritical Ising fusion category},
\end{split}
\end{equation}
and propose that these symmetries govern global structures over the SCFT moduli space.

These fusion categories play a role analogous to the $U(1)_R$ symmetry in $\mathcal{N}=2$ SCFTs. Their presence raises a natural question: what global structures do such non-invertible symmetries induce over moduli space?

The purpose of this paper is to propose and study such structures—categorified analogues of the Bagger–Witten line bundle—that arise in families of SCFTs with non-invertible symmetries. Concretely, we propose two related geometric objects:
\begin{itemize}
    \item A \textbf{bundle of fusion rings}, acting on vector bundles of closed string states;
    \item A \textbf{stack}\footnote{A stack is a categorified analogue of a bundle or a sheaf, in which fibers form categories rather than sets or modules.} \textbf{of fusion categories}, acting on families of D-brane categories over moduli space.
\end{itemize}
The latter is more refined, while the former may be viewed as a rigidification. These proposals generalize the familiar relation between invertible symmetries and Bagger–Witten line bundles to a broader setting involving non-invertible categorical symmetries.

\begin{figure}[h]
    \centering
    \includegraphics[width=6cm]{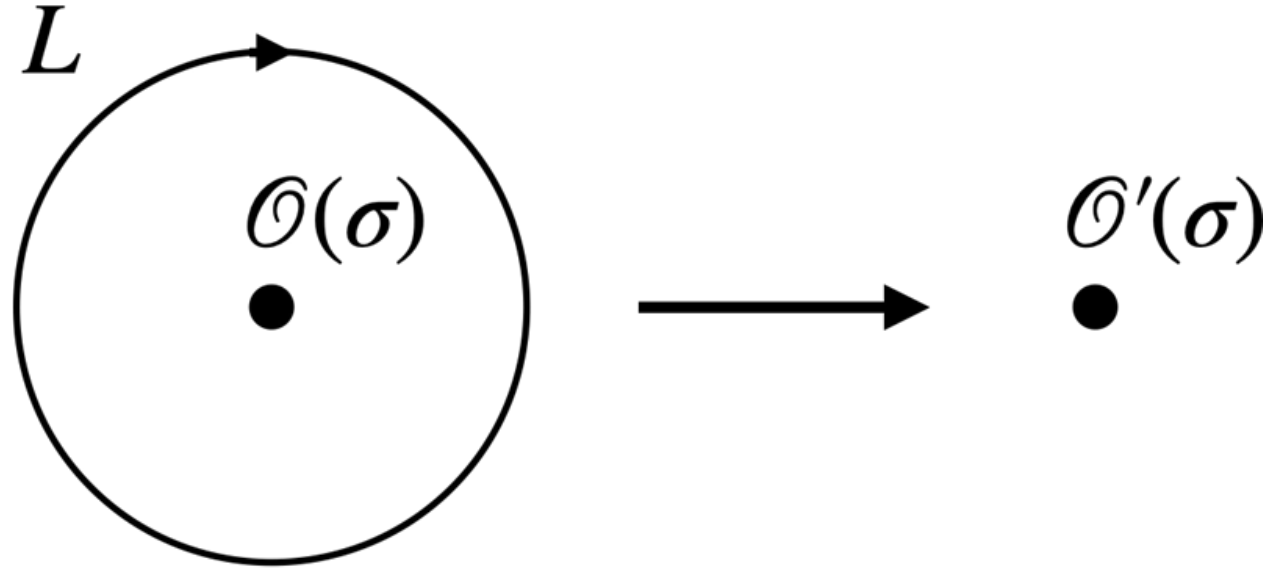}
    \caption{The action of a topological line operator on a local operator, associated to a vector bundle of the closed string state space over the moduli space.}
    \label{fig:bundle}
\end{figure}

\begin{figure}[h]
    \centering
    \includegraphics[width=14.5cm]{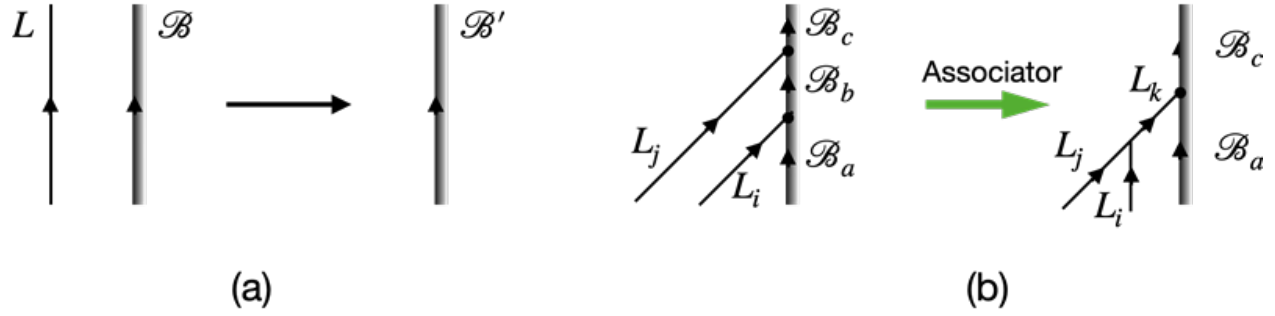}
    \caption{(a) The action of a topological line operator on a boundary, associated to a stack of D-brane categories over the moduli space. (b) The associator involving the symmetry category and the D-brane category implies that the global structure is a stack, rather than a bundle.}
    \label{fig:stack}
\end{figure}

We summarize their roles in Table~\ref{table:summ}, and illustrate their intuitive content in Figures~\ref{fig:bundle} and \ref{fig:stack}. In particular, Figure~\ref{fig:bundle} depicts how topological line operators, associated to elements of a fusion ring, act on closed string local operators via the state–operator correspondence. Figure~\ref{fig:stack}, in contrast, captures how fusion categories act on boundaries of the worldsheet, inducing categorical structure in the space of D-brane boundary conditions. The presence of nontrivial associators leads to the appearance of a stack, rather than a bundle.

\vspace{0.5em}

\vspace{0.5em}

\begin{table}[h]
\begin{tabular}{|c|c|c|}
\hline
&&\\
{\bf Symmetry} & {\bf Vector spaces of closed string states} & 
{\bf Categories of D-branes} \\ 
&& \\
\hline \hline
&&\\
Non-anomalous 
 & Vector bundle acted on by
& Stack acted on by 
\\invertible $G$ symmetry
& p-pal $G$-bundle $P$
& p-pal $G$-bundle $P$
\\
&&\\
\hline
&&\\ 
Anomalous  &
Vector bundle acted on by 
&
2-group bundle
\\
invertible $G$ symmetry,
& p-pal $G$-bundle $P$
& (stack of module cat's over
\\ 
$\alpha \in H^3(G,U(1))$
&(rigid. of Vec$(G,\alpha)$-stack for $G$ finite) & Vec$(G,\alpha)$ for $G$ finite)\\
&&\\
 \hline
&&\\
Non-invertible &
vector bundle acted on by 
& stack of module cat's
\\
fusion cat. ${\cal C}$ symmetry & fusion ring bundle (rigid. of $\mathcal{C}$-stack) 
&  over fusion category ${\cal C}$\\
&&\\
\hline
\end{tabular}
\label{table:summ}
\caption{
Symmetries and proposed interpretation of fibering closed string states and D-branes over the moduli space.
For invertible $G$ symmetries, fiberings of all bundles and stacks are governed by a single principal $G$-bundle.  For non-invertible symmetries, fiberings of all bundles and stacks are governed by a single stack of fusion categories.  The case of invertible symmetries with a 't Hooft anomaly forms a nontrivial bridge between the limiting cases.}
\end{table}

We now turn to the question of how these moduli space structures relate to spacetime physics. While our primary focus lies in the worldsheet SCFT perspective, it is often illuminating to consider how global symmetries manifest in the low-energy effective theory. In many familiar cases, worldsheet global symmetries give rise to gauge fields in spacetime, and the resulting geometric structures—such as line bundles or stacks—can be viewed as moduli space shadows of those gauge symmetries.

In all three cases we consider (as listed in Table \ref{table:summ})—non-anomalous invertible symmetries, anomalous invertible symmetries, and non-invertible symmetries—we propose corresponding structures over the SCFT moduli space. For non-anomalous invertible symmetries, the expected principal $G$-bundle acts on the fibering of both closed string Hilbert spaces and D-brane categories. For anomalous invertible symmetries, we show that the moduli space structure is similarly acted uopn by a bundle of fusion categories Vec$(G,\alpha)$, and that this structure matches a known gauge-theoretic implementation in terms of 2-group symmetries in spacetime.\footnote{See Appendix~\ref{app:fusion-2gp} for a discussion of this equivalence.}

For non-invertible symmetries—such as those arising from Ising or tricritical Ising categories—we again propose a moduli space structure: a stack of fusion categories fibered over the family of SCFTs. However, the nature of the corresponding spacetime symmetry remains obscure. While it is expected that these non-invertible symmetries lift to higher-categorical structures\footnote{At leading order in string coupling.}, a concrete gauge-theoretic realization is currently lacking.

We emphasize, therefore, that while spacetime perspectives offer valuable physical intuition, the moduli space structures we propose stand on their own. In the anomalous invertible case, the agreement with known spacetime realizations serves as evidence that these structures capture genuine physical data. This motivates extending the same geometric framework to the non-invertible case, even though the associated spacetime interpretation remains to be fully understood.

\vspace{0.5em}

The remainder of this paper is organized as follows. Section~\ref{sect:genl:nonanom} reviews the Bagger–Witten bundle from the moduli space perspective and motivates its generalization to fusion-ring and fusion-category structures. Section~\ref{sect:genl:anom} analyzes the intermediate case of anomalous invertible symmetries and introduces fusion-category descriptions of Vec$(G,\alpha)$-bundles. Section~\ref{sect:noninv} turns to non-invertible symmetries and formulates general conjectures about the resulting stack-like structures over moduli space. Section~\ref{sect:exchol} focuses on exceptional holonomy, proposing that moduli spaces of $G_2$ and $\mathrm{Spin}(7)$ SCFTs carry stacks of fusion categories related to (tricritical) Ising models. We describe how these categories act on topological observables and D-brane sectors.

We also include Appendix~\ref{app:fusion-2gp}, which clarifies the relation between 2-group symmetries and the fusion categories Vec$(G,\alpha)$ appearing in the anomalous case.

Explicitly constructing Bagger–Witten line bundles over Calabi–Yau moduli spaces is already a subtle problem. We expect that constructing the proposed categorified structures over exceptional holonomy moduli spaces will be even more intricate. Accordingly, we leave such constructions for future work.

\section{Non-anomalous invertible symmetries}   \label{sect:genl:nonanom}

In this section, we briefly review the appearance of bundles over moduli spaces of SCFTs, whose structure groups are determined by global symmetries of those SCFTs. Our focus is on the Bagger–Witten line bundle and its associated worldsheet $U(1)_R$ symmetry, as it is closely related to our later generalization for the $G_2$ and Spin(7) cases in Section~\ref{sect:exchol}. We also give brief remarks on examples involving finite invertible symmetries in A model topological strings and chiral symmetries in heterotic string compactifications.

As mentioned in Section \ref{sec:intro}, whenever one has a family of CFTs with an (ordinary) global symmetry $G$, one naturally gets a $G$ bundle over the moduli space, whose transition functions encode how the CFTs are related across
coordinate patches on the moduli space. Essentially, one must relate the operators and so forth to those
on the other, and the only possible relation is via a global symmetry, hence one has group elements on overlaps.  (For a bundle on a manifold, transition functions must close to the identity on triple overlaps; however, that condition is weakened for bundles interpreted as living on stacks.  As a result, we will largely omit discussion of the triple overlap condition, as one can shift from a moduli space to a gerbe over it.)

We can also understand this structure from the point of view of the spacetime effective theory. In string compactifications, non-anomalous worldsheet global symmetries imply gauge symmetries in spacetime. 
A large open patch of the SCFT moduli space can be identified in supergravity with the Higgs moduli space of scalar vevs in the
lower-dimensional effective field theory, and transition functions of the (restriction of the) bundle on the moduli space then reflect gauge transformations in the spacetime gauge theory.

In principle this dictionary is not exclusive to zero-form symmetries, or continuous groups.
The paper \cite{Pantev:2024kva} studied two-dimensional A-model open string theories
realizing Chern-Simons theories as the target-space string field theories,
including for possibly non-simply-connected gauge groups in the target-space
Chern-Simons theory.  For example, it distinguished $SU(2)$ Chern-Simons from
$SO(3)$ Chern-Simons.  The distinction revolved around the presence of a 
non-anomalous global worldsheet ${\mathbb Z}_2$ symmetry, which in the target space
becomes a gauged ${\mathbb Z}_2$ one-form symmetry, changing the $SU(2)$ gauge group to
$SO(3)$ by gauging its center.

\subsection{$U(1)_R$-induced structures on Calabi–Yau SCFT moduli spaces}

Let us now turn to the moduli space structures induced from $U(1)_R$ symmetries of Calabi-Yau SCFTs, including Bagger-Witten bundles and stack of D-brane categories, in more detail.

We start with a lighting overview of Bagger-Witten line bundles.  They were originally introduced in the context of four-dimensional $\mathcal{N}=1$ supergravity in \cite{Witten:1982hu} and then were interpreted within worldsheet SCFTs in \cite{Periwal:1989mx,Distler:1992gi}. For a review of recent developments, see \cite{Sharpe:2024rwb}.

Bagger-Witten line bundles arise over moduli spaces of 2D $\mathcal{N}=2$ chiral SCFTs, due to the fact that the theories possess a global $U(1)_R$ symmetry. The worldsheet $U(1)_R$ symmetry is generated by the spin-1 current $J(z)$ in the $\mathcal{N}=2$ superconformal algebra:
\begin{subequations}\label{eq:N=2 SCA}
\begin{align}
T(z) T(w) &\sim \frac{c/2}{(z-w)^4} + \frac{2T(w)}{(z-w)^2} + \frac{\partial T(w)}{z-w}, \\
T(z) J(w) &\sim \frac{J(w)}{(z-w)^2} + \frac{\partial J(w)}{z-w}, \\
T(z) G^\pm(w) &\sim \frac{3}{2} \frac{G^\pm(w)}{(z-w)^2} + \frac{\partial G^\pm(w)}{z-w}, \\
J(z) J(w) &\sim \frac{c/3}{(z-w)^2}, \\
J(z) G^\pm(w) &\sim \pm \frac{G^\pm(w)}{z-w}, \\
G^+(z) G^-(w) &\sim \frac{2c/3}{(z-w)^3} + \frac{2 J(w)}{(z-w)^2} + \frac{2 T(w) + \partial J(w)}{z-w}, \\
G^\pm(z) G^\pm(w) &\sim 0.
\end{align}
\end{subequations}
where $T(z)$ is the stress-energy tensor and $G^{\pm}$ are the two supercurrents.
Applying the $U(1)$ current, one can construct spectral flow operators ${\cal U}_{\theta}$,
of which ${\cal U}_{\pm 1/2}$ exchange NS and R sectors and so are responsible on the worldsheet for the implementation of spacetime supersymmetry.

An alternative viewpoint comes from the reduction of the structure group passing from K\"ahler ($U(d)$ holonomy) to Calabi-Yau ($SU(d)$ holonomy):
\begin{equation}\label{eq: u(1) from reduction}
    U(1)\cong \frac{U(d)}{SU(d)}.
\end{equation}
This reduction of structure group $U(d) \to SU(d)$ corresponds, on the worldsheet, to the presence of a global $U(1)_R$ symmetry in the SCFT. The $U(1)$ factor in $U(d)$ acts as a rotation of the worldsheet fermions, and becomes the generator of the $U(1)_R$ current in the $\mathcal{N}=2$ superconformal algebra. As a result, this $U(1)_R$ symmetry can be interpreted as the worldsheet shadow of the residual $U(1)$ geometric structure. In particular, the spectral flow operator $\mathcal{U}_{1/2}$ is naturally interpreted as twisting worldsheet fermions by the $U(1)$ in $U(d)$. This accounts for the name “spectral flow”: it realizes the spectral flow automorphism in the SCFT, exchanging different spin structures (i.e., NS $\leftrightarrow$ R sectors) in a way that depends on the degree of the canonical bundle — that is, the geometry of the Calabi–Yau. While this interpretation may seem redundant for Calabi–Yau compactifications, it will prove essential in the exceptional holonomy cases discussed in Section~\ref{sect:exchol}, where the global structure group data becomes nontrivial over moduli space.

The spectral flow operators ${\cal U}_{\pm 1/2}$ couple to the Bagger-Witten line bundles over the moduli space of SCFTs,
and the square ${\cal U}_1$ couples to the Hodge line bundle of holomorphic top-forms $\Omega^{(d,0)}$ on the Calabi-Yau, over the moduli space of complex structures. (Each should be interpreted as line bundles associated to an underlying principal $U(1)$ bundle, ultimately defined by the SCFT and its moduli space.)  From the perspective of calibrated geometry (see \cite{joyce2007riemannian} for a pedagogical introduction), the holomorphic top-form serves as the calibrated form characterizing Calabi–Yau manifolds, but not generic K\"ahler manifolds. Its existence implies the presence of covariantly constant spinors -- a hallmark of $SU(d)$ holonomy -- and underlies the preservation of spacetime supersymmetry in Calabi–Yau compactifications. 

From the worldsheet SCFT perspective, a section of the Hodge line bundle can be explicitly constructed from a distinguished operator associated to the holomorphic top-form $\Omega^{(d,0)}$:
\begin{equation}
 I= \Omega^{(d,0)}_{i_1\cdots i_d}\psi_-^{i_1}\cdots \psi_-^{i_d}.
\end{equation}
It was shown in \cite{Odake:1988bh} that this operator is a chiral primary\footnote{In $\mathcal{N}=2$ superconformal algebra, a chiral primary means it is a primary operator w.r.t $T(z)$ and has a trivial OPE with $G^+(z)$ (thus chiral).}, thus corresponds to a state in the closed string state space under the state/operator correspondence. One crucial property of this operator is that it carries $d$-charge under the $U(1)_R$ symmetry:
\begin{equation}
    J(z)I(w)\sim \frac{d}{z-w}I(w).
\end{equation}
Therefore, for the principal $U(1)_R$ bundle over the Calabi-Yau SCFT moduli space, $I(w)$ transforms accordingly as $I(w)\rightarrow e^{i\alpha d}I(w)$, thus serves as a section for the associated line bundle\footnote{$I(z)$ together with $T(z), G^{\pm}(z)$ and $J(z)$ generate an extension of the $\mathcal{N}=2$ superconformal algebra \cite{Odake:1988bh}.}. Although $I(w)$ is sometimes referred to as a spectral flow operator in the literature, we distinguish it from the canonical spectral flow operator $\mathcal{U}_{\pm 1/2}$ that exchanges NS and R sectors via a $U(1)_R$ rotation. Rather, $I(w)$ corresponds to $\mathcal{U}_1$, implementing a full $2\pi d$ rotation in $U(1)_R$, and geometrically reflects the global section of the canonical bundle $\Omega^{(d,0)}$ via worldsheet fermions.

A more mathematical understanding of Bagger-Witten line bundles over moduli spaces of Calabi-Yau's was proposed in
\cite{Donagi:2017mhd,Donagi:2019jic} (see \cite{Gu:2016mxp} for the special case of elliptic
curves).
From for example the fact that they can be understood as bundles of spectral flow operators $\mathcal{U}_{\pm 1/2}$ in the 
worldsheet $N=2$ SCFT, Bagger-Witten line bundles were geometrically interpreted as bundles of covariantly constant spinors over the moduli space of complex structures, in the same way that the Hodge line bundle is a bundle of nowhere-zero holomorphic volume forms over the moduli space of complex structures (i.e.~the relative canonical bundle). In fact, the tensor square\footnote{For the reader not familiar with the notion of square/square root of line bundles, think of the spin structure for a 2D Riemann surface, which can be regarded as a line bundle whose tensor square is just the canonical bundle $K=T^*\Sigma^{1,0}$ (bundle of holomorphic 1-forms).} of a Bagger-Witten line bundle is the Hodge line bundle: $\mathcal{L}_{\text{BW}}^{\otimes 2}\cong \mathcal{L}_{\text{H}}$. In particular, their first Chern classes satisfy $c_1(\mathcal{L}_{\text{H}})= 2 c_1(\mathcal{L}_{\text{BW}})$.  

Now, on a Calabi-Yau 3-fold, for example, there are two different covariantly constant spinors, which from the definition above means that there are two different Bagger-Witten line bundles over the corresponding moduli space.
Consistency of that pair of bundles with the original picture of \cite{Witten:1982hu} is tied to flatness\footnote{
Bagger-Witten line bundles for moduli spaces of Calabi-Yau's are now believed to be flat (but topologically/globally nontrivial), both from mathematics arguments over moduli spaces
of Calabi-Yau manifolds \cite{todorov1,todorov2} and from physics arguments concerning moduli spaces of SCFTs \cite{Gomis:2015yaa}.  Explicit examples have also recently been constructed \cite{Gu:2016mxp,Donagi:2017mhd,Donagi:2019jic}, demonstrating that the Bagger-Witten line bundle is flat but also nontrivial.  (See also \cite{Donagi:2017vwh} for related computations in higher-dimensional examples.)
} of the Bagger-Witten line bundle, as discussed in \cite{Donagi:2019jic}.

In passing, sometimes instead of discussing a bundle over a moduli space ${\cal M}$, it is more natural to discuss a bundle over a universal family.  For example, in the case of the Hodge line bundle, for ${\cal M}$ the moduli space of complex structures with Hodge
line bundle $L_\text{H} \rightarrow {\cal M}$, we could also discuss the universal family
$\pi: {\cal T} \rightarrow {\cal M}$, whose fibers are the Calabi-Yau's themselves, and carrying a bundle $\pi^* L_H$.  In various examples, we will sometimes use such an alternative description.  See for example \cite[section 3.2.1]{Donagi:2019jic} for more information in the case of Hodge line bundles.

\subsubsection{Bundles of closed string states}
Let us take a closer look at the bundles of closed string states under the $U(1)_R$ symmetries in $\mathcal{N}=(2,2)$ Calabi-Yau SCFTs. In these unitary $(2,2)$ SCFTs, the left- and right-moving sectors each contain the $\mathcal{N}=2$ superconformal algebra in Eq. (\ref{eq:N=2 SCA}), and with it a corresponding $U(1)$ R-current. These two currents, denoted $J$ and $\bar{J}$, are the building blocks for two distinguished combinations of symmetries that act on the full closed string Hilbert space:
\begin{equation}
    U(1)_V: J + \bar{J}, \qquad U(1)_A: J - \bar{J}.
\end{equation}
These combinations play a central role in the structure of moduli spaces.

The moduli space of Calabi-Yau SCFTs is locally parametrized by marginal operators preserving supersymmetry. These operators correspond to states in the closed string spectrum saturating the BPS condition $|q| = 2h = \frac{c}{3}$, where $q$ is the $U(1)$ R-charge and $h$ the conformal weight. They are left- and right-chiral (denoted as $c$ with $q>0$) or anti-chiral (denoted as $a$ with $q<0$) in various combinations. The precise transformation properties of these states under $U(1)_V$ and $U(1)_A$ provide a natural classification:
\begin{itemize}
    \item $(c,c)$ states, which are chiral in both sectors, carry $U(1)_V$ charge and are neutral under $U(1)_A$;
    \item $(c,a)$ states, which are chiral on the left and anti-chiral on the right, are neutral under $U(1)_V$ but carry $U(1)_A$ charge.
\end{itemize}
as well as $(a,a)$ and $(a,c)$ states with opposite charges accordingly.

This distinction is not merely formal: it reflects the geometric nature of the corresponding moduli. For Calabi–Yau, the $(c,c)$ primaries correspond to infinitesimal deformations of complex structure, while the $(c,a)$ primaries encode K\"{a}hler moduli.

To be more concrete, consider SCFTs arising from Calabi–Yau 3-folds. In such models, the closed string state space includes ground states labeled by Hodge numbers of the target\footnote{see e.g. \cite{Greene:1996cy} for a review of the role of chiral rings and moduli in Calabi–Yau compactifications}:
\begin{itemize}
    \item The $(c,c)$ ring corresponds to elements in $H^{2,1}(X)$ and $H^{3,0}(X)$; the former parametrize complex structure deformations, while the latter corresponds to the holomorphic top-form $\Omega^{(3,0)}$.
    \item The $(c,a)$ ring corresponds to elements in $H^{1,1}(X)$; these define variations of the K\"{a}hler form and hence of the K\"{a}hler structure moduli.
\end{itemize}

These sectors are not just algebraic artifacts—they fiber nontrivially over the moduli space $\mathcal{M}$. For example, the $(c,c)$ state space define a fiber of a holomorphic bundle $\mathcal{L}_C$ over the complex structure moduli space, equipped with a natural connection arising from the operator product algebra of the chiral ring. A $(c,c)$ primary $O$ with $(q,\bar{q})$ $U(1)$ R-charges corresponds to a section of this line bundle, whose gluing condition along the moduli space parameterized by $p$ is given by the $U(1)_V$ transformation
\begin{equation}
    O_p \rightarrow O_{p'}e^{i\alpha_{pp'} (q+\bar{q})}
\end{equation}
In the simplest case, the ground state in the $(c,c)$ sector—associated to the holomorphic top-form $\Omega^{(3,0)}$—generates a holomorphic line bundle. This is the exactly the Bagger–Witten line bundle associated to the principal $U(1)_V$ bundle. It reflects how the choice of vacuum varies as one moves across the moduli space. Likewise, the $(c,a)$ states define a line bundle $\mathcal{L}_K$ associated to the principal $U(1)_A$ bundle over the K\"{a}hler moduli space. The section of this line bundle transformed under $U(1)_A$ symmetry as
\begin{equation}
    O_p \rightarrow O_{p'}e^{i\alpha_{pp'} (q-\bar{q})}.
\end{equation}

Mirror symmetry exchanges these structures. For a mirror pair $(X,Y)$ of Calabi–Yau $3$-folds, the roles of $(c,c)$ and $(c,a)$ sectors are interchanged. In particular, the complex structure moduli space of $X$ is isomorphic to the K\"{a}hler moduli space of $Y$, and vice versa. This duality maps the $U(1)_V$ symmetry of one theory to the $U(1)_A$ symmetry of the mirror. As a result, the bundle structure over the moduli space transforms accordingly: the Bagger–Witten line bundle associated to the $(c,c)$ ground state on $X$ corresponds, under mirror symmetry, to the analogous structure over the K\"{a}hler moduli space of $Y$, encoded in the $(c,a)$ sector\footnote{Strictly speaking, to make this mirror map work, some `corners' of the moduli space may need to be omitted.}.

Thus, even in this comparatively familiar setting, the relation between global symmetries and fiber structures over moduli spaces is both geometrically rich and physically meaningful. The structure of the closed string Hilbert space—organized according to its $U(1)_V$ and $U(1)_A$ charges—already reveals the necessity of line bundles over the moduli space.

\subsubsection{Stacks of D-brane categories}

The closed string sector of a $\mathcal{N}=(2,2)$ superconformal field theory defines vector bundles and line bundles over the moduli space $\mathcal{M}$, with fibers corresponding to chiral primaries, and connections given by variations of the background. In contrast, the open-string sector encodes a richer structure: that of a stack of categories, reflecting not only how branes vary over moduli space, but how the entire structure of morphisms and their composition laws change globally.

Let us begin with basic definitions. A D-brane in such a theory is characterized by a conformally invariant boundary condition, preserving half of the supersymmetry. A-branes preserve a linear combination $Q_L + \bar{Q}_R$ of the left- and right-moving supercharges, while B-branes preserve $Q_L + Q_R$. These boundary conditions are defined in terms of the $\mathcal{N}=2$ superconformal algebra, and are naturally associated with distinct R-symmetry structures:
\begin{itemize}
    \item A-brane boundary states are graded by $U(1)_V$ charge;
    \item B-brane boundary states are graded by $U(1)_A$ charge.
\end{itemize}

Now consider the open string Hilbert space $\mathcal{H}_{E,F}$ between two branes $E$ and $F$\footnote{In target space, $E$ and $F$ are vector bundles over submanifolds of the Calabi-Yau.}. This space decomposes into sectors labeled by the $U(1)$ charge $q$
\begin{equation}
    \mathcal{H}_{E,F} = \bigoplus_q \mathcal{H}^q_{E,F}, \quad \text{with}~q \in \mathbb{Z},
\end{equation}
where $q=q_V$ for A-branes and $q=q_A$ for B-branes. Under the state-operator correspondence, these open string states correspond to boundary-changing operators between $E$ and $F$ -- and their OPEs define composition maps
\begin{equation}
   \mathcal{H}^q_{E,F} \otimes \mathcal{H}^{q’}_{F,G} \to \mathcal{H}^{q+q’}_{E,G}.
\end{equation}

The existence of worldsheet boundaries, boundary-changing operators and the associativity of their compositions leads to the following categorical structure\footnote{
    See e.g.~\cite{Sharpe:1999qz,Fuchs:2001qc,Aspinwall:2001pu,Sharpe:2003dr,Aspinwall:2004jr} for discussions of categories of branes.}:
\begin{itemize}
    \item Objects: D-branes as worldsheet boundaries;
    \item Morphisms: Boundary-changing operators;
    \item Composition: OPEs of boundary-changing operators.
\end{itemize}

\begin{figure}[h]
    \centering
    \includegraphics[width=9cm]{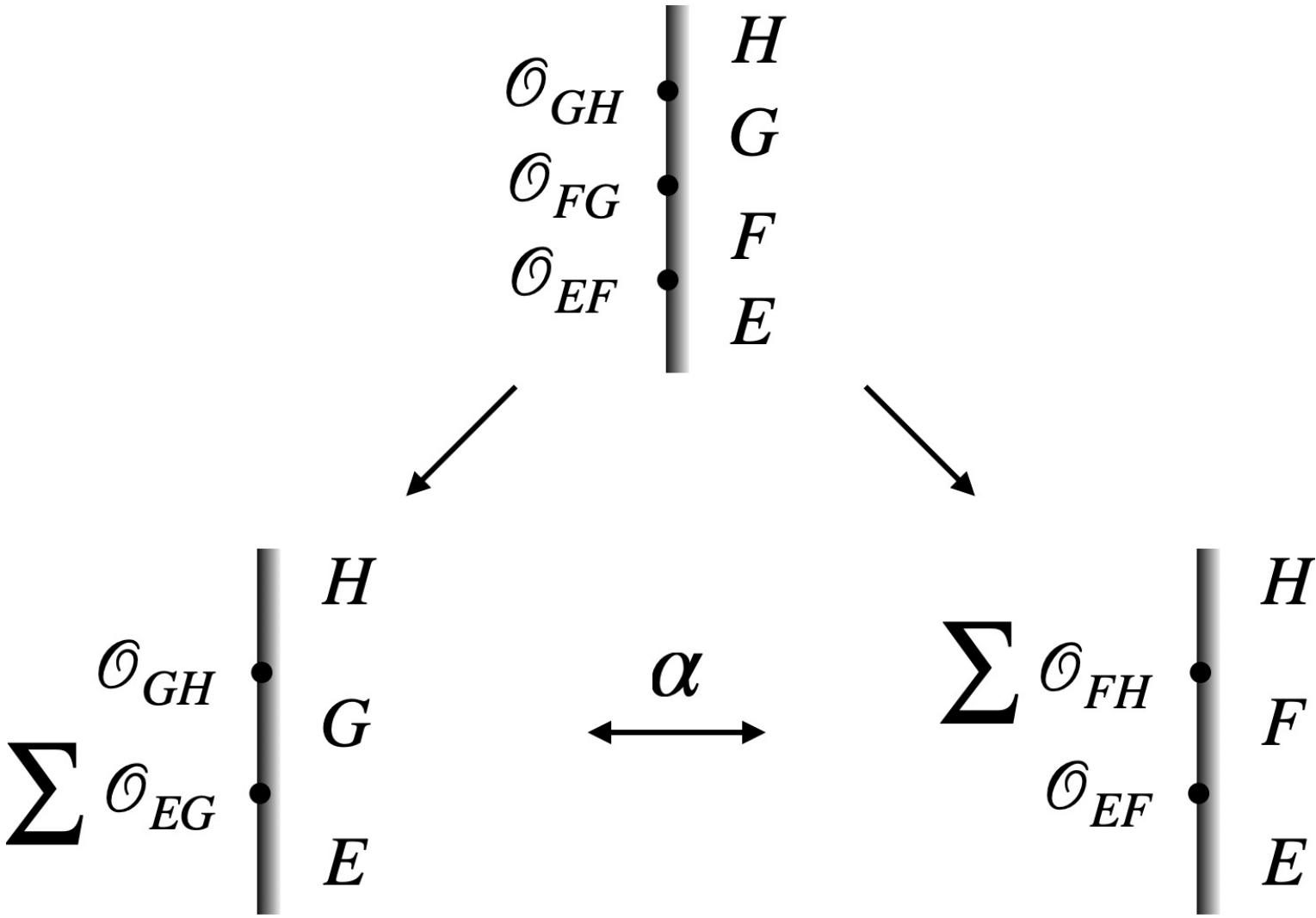}
    \caption{The associator of morphisms in D-brane categories. $E, \cdots H$ are worldsheet boundary components as D-branes, i.e. objects in the category, while $\mathcal{O}_{EF},\cdots$ are boundary changing operators, i.e., morphisms in the category. The associator labels isomorphisms between OPEs of boundary changing operators. Under state-operator correspondence, $\mathcal{O}_{EF}$ corresponds to a state in the open string Hibert space $\mathcal{H}^q_{E,F}$, where $q$ is the $U(1)$ R-charge carried by $\mathcal{O}_{EF}$. Altogether, this categorical structure is fibered over the moduli space of SCFTs leads to a stack, associated to principal $U(1)$ bundles, under which morphisms in the fibered category are glued.}
    \label{fig:bdyasso}
\end{figure}

Importantly, the composition in D-brane categories is not strictly associative at the level of operators; rather, it is associative up to homotopy. The homotopy associativity is captured by the presence of an associator, a higher structure map which encodes how the three-point boundary OPEs satisfy associativity only up to an exact term. See Figure \ref{fig:bdyasso} for an intuitive depiction. Mathematically, this data is precisely what is needed for an $A_\infty$ category.

This higher associativity structure distinguishes a stack of categories from a vector bundle\footnote{More precisely, a stack of categories is a sheaf of categories satisfying descent and effective gluing conditions, generalizing the notion of a sheaf of vector spaces.}. A bundle of categories over moduli space would assume a fixed composition rule fiberwise, but the associator—along with possible autoequivalences induced by monodromies—requires a more flexible structure: a \emph{stack} fibered in $A_\infty$ categories. The fiber at each point $p$ in moduli space $\mathcal{M}$ is a brane category, but the transition functions between patches involve not just isomorphisms of categories, but equivalences preserving the $A_\infty$ structure up to coherent higher morphisms.

The variation of complex structure moduli or K\"{a}hler moduli induces changes not only in the set of objects and morphisms, but in the higher composition maps themselves. 
Just as chiral primaries in the closed string state space furnish sections of line bundles over $\mathcal{M}$ associated to $U(1)$ symmetries, the D-brane categories furnish a stack over $\mathcal{M}$, graded by $U(1)$ charges and equipped with higher composition data.

As in the closed string case, mirror symmetry exchanges these two sectors: for a mirror pair Calabi-Yau $(X, Y)$, one has
\begin{equation}
    \text{A-brane} \leftrightarrow \text{B-brane}, \qquad U(1)_V \leftrightarrow U(1)_A,
\end{equation}
and hence the stack of B-brane categories on $Y$ corresponds to the stack of A-brane categories on $X$\footnote{Here we use $\leftrightarrow$ instead of $\simeq$ to label a very rough correspondence between the two categories, due to the subtlety that there are coisotropic A-branes which may not be encoded by $\mathcal{F}(X)$.}. This duality exchanges the roles of the two $U(1)$ symmetries and their associated twistings.

More formally, the underlying categories for (Lagrangian) A-branes and B-branes are well-known as:
\begin{itemize}
    \item (Lagrangian) A-brane category: Fukaya category\footnote{Strictly speaking, generic A-branes are associated to coisotropic submanifolds of the target Calabi-Yau, which are beyond Fukaya categories.} $\mathcal{F}(X)$ with objects as Lagrangian submanifolds $L\subset X$ and morphisms as Floer (co)homology groups ${\rm CF}^q(E, F)$.
    \item B-brane category: Derived category  $\mathcal{D}^b(X)$ with objects as complexes of coherent sheaves on X, and morphisms as Ext groups ${\rm Ext}^q(E,F)$.
\end{itemize}
While we expect that D-brane stacks acted by $U(1)_V$ and $U(1)_A$ over moduli spaces are rigorously described via these category languages, we leave the precise investigation for a future work.

To summarize, the D-brane category organize into a stack of $A_\infty$ categories graded by $U(1)$ R-charges, whose structure reflects the global variation of brane and open string data, the global $U(1)$ information, and the nontrivial associators arising from OPEs of boundary-changing operators.

\subsubsection{Comments on $\mathcal{N}=(0,2)$ models}
In principle, there are other bundles over moduli spaces of SCFTs that arise similarly.
For example, consider a moduli space of two-dimensional $\mathcal{N}=(0,2)$ supersymmetric theories describing a perturbative heterotic string compactification.  Typically, these theories have a non-anomalous chiral left-moving global $U(1)$ symmetry, which on the $\mathcal{N}=(2,2)$ locus becomes the left $R$-symmetry.  Off the $\mathcal{N}=(2,2)$ locus, it still exists, and forms part of the worldsheet symmetry needed to build the spacetime gauge symmetry.
More to the point for our purposes, over the moduli space of $\mathcal{N}=(0,2)$ SCFTs, there is a corresponding line bundle, because of that left-moving $U(1)$ symmetry.  This induces an analogue of the Bagger-Witten line bundle, but for simplicity let us discuss its tensor square, which can be understood mathematically as a line bundle of trivializations $\det E \rightarrow {\cal O}$, where ${\cal E}$ is the heterotic gauge bundle (taken to have trivial determinant).  This is an analogue of the Hodge line bundle, and on the $\mathcal{N}=(2,2)$ locus, it becomes precisely that.
This is discussed further in e.g.~\cite[section 6]{Gu:2016mxp}.

\section{Anomalous invertible symmetries}  \label{sect:genl:anom}

Our primary goal is to understand categorified structures over SCFT moduli spaces arising from non-invertible symmetries—particularly those appearing in compactifications on exceptional holonomy manifolds. As we will identify in Section \ref{sect:exchol}, these symmetries are Ising fusion category and tricritical Ising fusion category. From a physical perspective, these symmetries share certain features with 't Hooft anomalous invertible symmetries — obstruction to gauging — and are easier to understand in established language. 
However, to our knowledge, the implications of anomalies of worldsheet global symmetries for moduli space structures (by which we mean bundles, stacks, or higher-categorical structures informed by symmetry data) seem to remain largely unexplored in the physics literature. This motivates us to study the anomalous invertible case first as a controlled setting.

As a result, in this section we focus on invertible global symmetries with a 't Hooft anomaly\footnote{Unless otherwise specified, we use the term ‘anomaly’ to refer to ’t Hooft anomalies, not chiral anomalies, which explicitly break the symmetry.}, which serve as a stepping stone toward the non-invertible case.

\paragraph{Chiral anomalies.} Before addressing 't Hooft anomalies directly, we briefly comment on a closely related but conceptually distinct situation — global symmetries with \emph{chiral anomalies}. The key distinction is whether the global symmetry remains exact in the quantum theory. In chiral anomalies (e.g., the 4D Adler–Bell–Jackiw anomaly \cite{Adler:1969gk, Bell:1969ts}), the symmetry current is not conserved, and the symmetry does not survive quantization. In contrast, a 't Hooft anomaly preserves the symmetry at the quantum level, but obstructs its gauging, providing a powerful constraint via anomaly matching \cite{tHooft:1979rat}.

In string theory, the connection between worldsheet global symmetries and spacetime gauge symmetries hinges on the construction of vertex operators for spacetime gauge bosons. These operators rely on conserved currents for worldsheet global symmetries. In the presence of a chiral anomaly, the current is non-conserved, and the corresponding vertex operator fails to be BRST-closed — hence, no spacetime gauge field arises.

This expectation is borne out in concrete examples. Consider a heterotic compactification with gauge bundle $L \oplus L^{-1}$, where $L$ is a line bundle of non-zero first Chern class. Such configurations arise at stability walls as solutions to the Donaldson–Uhlenbeck–Yau equations (see, e.g., \cite{Sharpe:1998zu, Anderson:2009sw, Anderson:2009nt}). These configurations offer a clear instance where a worldsheet chiral anomaly corresponds to a massive $U(1)$ gauge field in spacetime. The left-moving fermions in the $L$ summand transform under a global worldsheet $U(1)$ symmetry, which is anomalous. This has been studied extensively (see \cite{Dine:1987xk,Honecker:2006dt,Honecker:2006qz,Donagi:2009ra,Anderson:2011ty,Harvey:2025wxy}; also \cite[Section 3.1]{Melnikov:2012cv}). In spacetime, the associated $U(1)$ gauge symmetry is Higgsed: the gauge-transformed parts of the $B$-field act as St\"uckelberg fields, rendering the $U(1)$ gauge boson massive.  In short:
\begin{equation*}
\text{worldsheet $U(1)$ with chiral anomaly} \quad \longleftrightarrow \quad \text{spacetime $U(1)$ Higgsed}.
\end{equation*}

\paragraph{'t Hooft anomalies.} Let us now turn to 't Hooft anomalies on the worldsheet. Suppose we have an invertible global symmetry group $G$ with 't Hooft anomaly classified by a cohomology class $\alpha \in H^3(G, U(1))$. When $G$ is finite, such an anomalous symmetry is best understood\footnote{Alternatively, it may be described in terms of 2-group symmetries; see Appendix~\ref{app:fusion-2gp}.} as a fusion category symmetry — specifically, $\text{Vec}(G, \alpha)$ — rather than as a group symmetry. Over the moduli space of SCFTs, we then expect structures determined by the Vec$(G,\alpha)$ symmetry:
\begin{itemize}
    \item a bundle of Vec$(G,\alpha)$ fusion rings\footnote{
    The fusion ring of any fusion category is generated by the simple objects of the fusion category, with multiplication given by the fusion category product \cite[section 4.9]{egno}. In this case,
    the fusion ring of Vec$(G,\alpha)$ matches that of Vec$(G)$ \cite[prop. 4.10.3]{egno}, so the anomaly is invisible at the level of the fusion ring.
    }, acting on e.g.~vector bundles of closed string states

    \item a stack of Vec$(G,\alpha)$ fusion categories, acting on e.g.~stacks of D-brane categories.
\end{itemize}
These structures serve as a model for what we expect in the non-invertible setting, where $\text{Vec}(G, \alpha)$ is replaced by the pertinent non-invertible fusion category.

Unlike chiral anomalies, 't Hooft anomalies do not break the global symmetry: the Noether current remains conserved, and one can still construct the associated BRST-closed vertex operator. As a result, the symmetry is still visible in the spacetime theory as a gauge symmetry. The anomaly obstructs gauging on the worldsheet, but does not eliminate the associated spacetime gauge field. The interesting question becomes: 

 \emph{What structure in the spacetime theory or over the SCFT moduli space reflects the presence of a worldsheet 't Hooft anomaly?}

To illustrate the structures we are interested in, we now turn to a concrete and illuminating example: the compact boson describing a toroidal compactification, in which the worldsheet theory exhibits a mixed ’t Hooft anomaly between momentum and winding symmetries. This model provides a minimal setting, where anomalies, their spacetime avatars, and moduli space structures can all be treated explicitly. See \cite{Harvey:2025wxy} for a recent complementary discussion.

In this example, the global symmetry group is $G = U(1)^2$, corresponding to conserved momentum and winding numbers. Although $G$ is not finite, it contains a diagonal $\mathbb{Z}_2$ subgroup which inherits a self-anomaly, and which admits a fusion category description. We will study this system from two complementary perspectives. In Section~\ref{sect:anom:spacetime}, we analyze the spacetime realization of the anomaly and its associated 2-group structure. In Section~\ref{sect:anom:modspace}, we examine the corresponding global structures over the moduli space.

\subsection{Spacetime implementation of anomalies}
\label{sect:anom:spacetime}

In order to demonstrate the idea, it is sufficient to consider bosonic string theory compactified on a circle $S^1$:
\begin{equation}
    M_{25}\times S^1.
\end{equation}
The subsector of the worldsheet theory describing this internal geometry $S^1$ is simply a $c=1$ compact boson. At generic radius, the worldsheet global symmetry takes the form:
\begin{equation}
\left( U(1)_m \times U(1)_w \right) \rtimes \mathbb{Z}_2,
\end{equation}
where $U(1)_m$ and $U(1)_w$ correspond to momentum and winding number symmetries, respectively.

An important property of this global symmetry is the mixed 't Hooft anomaly between the two $U(1)$ symmetries. To understand its manifestation in spacetime, we identify the associated spacetime gauge fields to $U(1)_m$ and $U(1)_w$. The string states corresponding to these symmetries are (see, e.g., \cite[Chapter 8]{Polchinski:1998rq}):
\begin{equation}
    \alpha_{-1}^\mu \tilde{\alpha}_{-1}^{25}+\alpha_{-1}^{25}\tilde{\alpha}_{-1}^\mu |0;k\rangle, ~~~\alpha_{-1}^\mu \tilde{\alpha}_{-1}^{25}-\alpha_{-1}^{25}\tilde{\alpha}_{-1}^\mu |0;k\rangle.
\end{equation}
The first state yields the Kaluza–Klein gauge boson from $A_\mu \sim g_{\mu 25}$, i.e., off-diagonal elements of the 26-dimensional metric, associated with $U(1)_m$. The second state gives rise to a gauge boson $B_\mu \sim B_{\mu 25}$, i.e., off-diagonal elements of the antisymmetric $B_{MN}$ ($M, N=0, 1,\cdots, 25$) field, associated with $U(1)_w$. To see this, consider the winding-state vertex operator:
\begin{equation}
    \partial X^{\mu} \bar{\partial}X^{25}-\partial X^{25}\bar{\partial}X^{\mu}=\bar{\partial}(X^{25}\partial X^{\mu})-\partial(X^{25}\bar{\partial}X^{\mu})
\end{equation}
not integrated to zero in the winding state since $X^{25}$ is not single-valued.
In this fashion we reproduce the vertex operators for the expected
spacetime $U(1)_m \times U(1)_w$ gauge symmetry, see e.g.~\cite{Narain:1986am}.

Having identified gauge fields $A_\mu$ and $B_\mu$ as spacetime counterparts for worldsheet $U(1)_m$ and $U(1)_w$ symmetries, the mixed 't Hooft anomaly can be seen as follows. The compactified 25-dimensional theory, in addition to the gravity sector, $A_\mu$ and $B_\mu$, has also $B_{\mu \nu}$ field, which comes from the 26-dimensional $B$-field. The key point is that in the 25D spacetime, the gauge transformation of $A_\mu$ induces a shift in the 2-form field $B_{\mu \nu}$, inherited from the 26D $B$-field \cite{Heidenreich:2020pkc}, 
\begin{equation}
    A^{(1)} \rightarrow A^{(1)} + d\chi, ~B^{(2)}\rightarrow B^{(2)}-\frac{1}{2\pi}B^{(1)}\wedge d\chi,
\end{equation}
where we suppress spacetime indices and use an upper index to denote the degree of the differential forms. This implies the well-defined 3-form flux in the 25-dimensional theory is 
\begin{equation}\label{eq:chernsimonslocal}
    \tilde{H}^{(3)}=dB_2+\frac{1}{2\pi}H^{(2)}\wedge A^{(1)},
\end{equation}
and hence a \emph{modified Bianchi identity}:
\begin{equation}\label{eq:fluxidentity}
     d\tilde{H}^{(3)}=H^{(2)}\wedge F^{(2)},
\end{equation}
where $\tilde{H}^{(3)}, H^{(2)}$ and $F^{(2)}$ are all globally well-defined, with $H^{(2)}$ being the curvature of the principal $U(1)$ bundle connection $B^{(1)}$. Therefore, the anomaly of the $U(1)_m\times U(1)_w$ symmetry on the worldsheet becomes a Green-Schwarz-like mechanism in spacetime.

To make contact with earlier discussions, in modern terms, we would say that this reflects a 2-group
$\tilde{G}$ symmetry\footnote{2-group symmetry in string theory was already recognized in \cite{Baez:2005sn} in the context of heterotic strings.}, where $\tilde{G}$ is an extension of $G = U(1) \times U(1)$ by the 2-group
$BU(1)$, with extension class determined by the worldsheet anomaly. This extension is a short exact sequence of 2-groups
\begin{equation}\label{eq:2gpsequence}
    1 \to BU(1) \to \tilde{G} \to U(1)\times U(1) \to 1, 
\end{equation}
classified topologically in cohomology\footnote{As explained in \cite{Schommer-Pries:2011vyj, Schreiber:2013pra} and summarized in \cite[section 3.3]{Kang:2023uvm}, the smooth cohomology refinement is computed not in terms of cochains $(U(1)\times U(1))^n \to U(1)$ \cite{Baez:2003yaq}, which gives a trivial cohomology group \cite{tonypriv}, but in terms of classes of maps of stacks.} by
\begin{eqnarray}
    H^3\left(U(1)\times U(1), U(1)\right) &\cong& H^4_{\rm sing}\left(B(U(1)\times U(1)),\Z\right),
    \\ &\cong& \left( H^4_{\sing}\left(B(U(1)),\Z\right)\right) \oplus \left( H^4_{\rm sing}\left(B(U(1)),\Z\right) \right)\\ & &  \oplus \, \left(H^2_{\sing}\left(B(U(1)),\Z\right)\otimes H_{\rm sing}^2\left(B(U(1)),\Z\right) \right) ,
    \\
    &\cong& \Z\oplus \Z \oplus \Z,
\end{eqnarray}
where $B(G)$ is the classifying space of the group $G$.

In purely topological terms, Equation~(\ref{eq:fluxidentity}) says that $H^{(2)}\wedge F^{(2)}$ has a trivial class in integral cohomology of $M$, meaning we have a homotopy-commutative diagram
\begin{equation}
    \begin{tikzcd}
                                                       &  & B(U(1)\times U(1))\cong B^2(\Z\times \Z) \arrow[dd, "c_2c_2"] \\
                                                       &  &                                                               \\
M \arrow[rr, "*"'] \arrow[rruu, "{(H^{(2)},F^{(2)})}"] &  & B^4\Z                                                        
\end{tikzcd},
\end{equation}
where $*:M\to B^4\Z$ is the homotopy-trivial map. By the fibration induced by the short exact sequence (\ref{eq:2gpsequence}), this means the $(H^{(2)},F^{(2)})$ fluxes come from a principal $\widetilde{G}$ (higher) bundle
\begin{equation}
 \begin{tikzcd}
                                                                                                     &  & B\widetilde{G} \arrow[dd]                                     \\
                                                                                                     &  &                                                               \\
                                                                                                     &  & B(U(1)\times U(1))\cong B^2(\Z\times \Z) \arrow[dd, "c_2c_2"] \\
                                                                                                     &  &                                                               \\
M \arrow[rr, "*"'] \arrow[rruu, "{(H^{(2)},F^{(2)})}" description] \arrow[rruuuu, dashed, bend left] &  & B^4\Z                                                        
\end{tikzcd}.
\end{equation}

We can furthermore pass to the smooth setting to see the identities at the level of connections. We have a twisted structure (in the sense of \cite{Sati:2009ic}) described by the following (smooth) homotopy-commutative diagram
\begin{equation}
   \begin{tikzcd}
                                                                           &  &                                                 &  & B_{\nabla}(U(1)\times U(1)) \arrow[dd, "dB\wedge A"] \arrow[lldd, "B_2", Rightarrow] \\
                                                                           &  &                                                 &  &                                                                                      \\
M \arrow[rrrruu, "{(A^{(1)},B^{(1)})}"] \arrow[rr, "\widetilde{H}^{(3)}"'] &  & {B^3_{\nabla,\text{triv}}U(1)} \arrow[rr, hook] &  & B^3_{\nabla}U(1)                                                                    
\end{tikzcd},\label{eq:commdiagramtwist}
\end{equation}
where $B_{\nabla}G$ is the smooth moduli stack of principal $G$-bundles with connection, and $B^3_{\nabla,\text{triv}}U(1)$ is the moduli stack of \textit{trivial} bundle 2-gerbes (principal $B^2U(1)$ higher bundles) with connection \cite{Schreiber:2013pra}. Here, $(A^{(1)},B^{(1)}):M\to B_{\nabla}(U(1)\times U(1))$ is a pair of connections $A^{(1)},B^{(1)}$ of principal $U(1)$ bundles, $\widetilde{H}^{(3)}:M\to B^3_{\nabla, \text{triv}}U(1) \hookrightarrow B^3_{\nabla}U(1) $ is a connection on a \textit{trivial} bundle 2-gerbe, and $dB\wedge A: B_{\nabla}(U(1)\times U(1))\to B_{\nabla}^3 U(1)$ is a universal Chern-Simons 2-gerbe. The diagram (\ref{eq:commdiagramtwist}) corresponds to the identity (\ref{eq:chernsimonslocal}), where $\widetilde{H}^{(3)}, B_2$, and hence the Chern-Simons 2-gerbe $dB^{(2)}\wedge A^{(1)} = H^{(2)}\wedge A^{(1)}$ on $M$, are globally-defined by the triviality of the bundle 2-gerbe underlying $\widetilde{H}^{(3)}$.

More geometrically, this mechanism is simply the result of performing a double dimensional reduction of a $B^3\Z$-cocycle (the $H$-flux) on the $S^1$ factor of a spacetime of the form $M\times S^1$. This reduction is given by the adjunction \cite{CC87,BMSS19}
\begin{equation}\label{eq:adjcyc}
    \text{Map}(M\times S^1,B^3\Z) \cong \text{Map}(M,[S^1\times B^3\Z]//S^1)=\text{Map}(M,\text{Cyc}(B^3\Z)),
\end{equation}
where $\text{Cyc}(B^3\Z)$ is the cyclic loop space of $B^3\Z$. A quick way to understand the Bianchi identities is to work at the level of rational homotopy theory \cite{GM13}. In this setting, the generators of the Sullivan model \cite{Sul73}  of $\text{Cyc}(B^3\Z)$ give a tractable statement of these identities. It is known \cite{VPB85} that the Sullivan model for $\text{Cyc}(B^3\Z)$ is generated by a pair of degree $2$ elements $c_2,\omega_2$ and one element $c_3$ of degree 3, satisfying the Bianchi identities $dc_2=d\omega_2=0$ and $dc_3 = \omega_2\wedge c_2$. At the rational level, then, a cocycle (namely, a map) on $M$ valued in $\text{Cyc}(B^3\Z)$ is described by differential forms $c_2,\omega_2\in \Omega^2(M), c_3\in\Omega^3(M)$, satisfying the identities above, namely $d_M c_2= d_M \omega_2=0$ and $d_Mc_3= c_2\wedge \omega_2$. This double-dimensional reduction matches the flux identities (\ref{eq:fluxidentity}).

To summarize so far, in a compactification on a circle, the worldsheet theory has a global $U(1)_m \times U(1)_w$ symmetry with a 't Hooft anomaly.
On the worldsheet, the global symmetry cannot be gauged.  Nevertheless, the global symmetry still defines a gauge symmetry in the spacetime theory, where the worldsheet 't Hooft anomaly manifests as a modified Bianchi identity, one which does not Higgs either $U(1)$ (unlike the case of a worldsheet chiral anomaly).

\paragraph{Anomalous $\mathbb{Z}_2$ subgroup and fusion category description.}
Although the full symmetry group $G = U(1)_m \times U(1)_w$ is continuous, it contains a diagonal $\mathbb{Z}_2$ subgroup generated by simultaneously inverting both momentum and winding numbers. This $\mathbb{Z}_2$ subgroup inherits a self-anomaly from the mixed anomaly of the ambient $U(1)^2$ symmetry\footnote{One way to understand this is to notice at the self-dual radius under T-duality, the corresponding $c=1$ CFT is the $\mathfrak{su}(2)_1$ WZW model, which has an anomalous $\mathbb{Z}_2$.}. As such, it provides a finite and tractable setting in which to examine the fusion category structure associated to anomalous symmetries.

In particular, the anomalous $\mathbb{Z}_2$ symmetry is described by a nontrivial element
$$
\omega \in H^3(\mathbb{Z}_2,U(1)) \cong \mathbb{Z}_2,
$$ 
leading to a fusion category symmetry $\mathrm{Vec}_{\mathbb{Z}_2}^\omega$. This fusion category provides a minimal example of a nontrivial associator structure, and will serve as a warm-up for more intricate fusion categories, such as the (tricritical) Ising categories that appear in exceptional holonomy compactifications.

From the spacetime point of view, this $\mathbb{Z}_2$ symmetry can be viewed as a finite subgroup of the $U(1)^2$ gauge symmetry. The corresponding gauge field is a discrete $\mathbb{Z}_2$ background field, and the anomaly again implies a 2-group structure. Specifically, the $\mathbb{Z}_2$ gauge theory is extended by a $BU(1)$ 2-group, with the $B$-field serving as the 2-form gauge field for the higher symmetry. This 2-group structure is straightforwardly inherited from the continuous 2-group $\tilde{G}$ previously described.

While this step—restricting to a subgroup—does not introduce new dynamics, it serves to highlight the appearance of a nontrivial fusion category symmetry in a finite and concrete setting. In particular, as we will see in Section~\ref{sect:anom:modspace}, it provides an explicit realization of a moduli space structure corresponding to $\mathrm{Vec}_{\mathbb{Z}_2}^\omega$, and helps motivate the more involved non-invertible symmetry structures over moduli space discussed in the next section.

\subsection{Structures over the moduli space}
\label{sect:anom:modspace}

We now turn to the structures over the CFT moduli space associated with the mixed 't Hooft anomaly discussed in the previous section. Our goal is to understand how the anomaly affects the fibering of closed and open string data over the moduli space, and how this leads to categorified geometric structures---specifically, vector bundles (governed by fusion rings) and stacks (governed by fusion categories).

We begin with the closed string sector, where the anomaly is invisible. We then turn to the D-brane sector, where the anomaly obstructs global equivariant structures and necessitates a stack-theoretic interpretation.

\subsubsection{Closed string states and principal bundles}
The compact boson theory compactified on a circle of radius $R$ possesses global $U(1)_m \times U(1)_w$ symmetries, corresponding to momentum and winding number conservation. The closed string Hilbert space consists of states $|n,w\rangle$ with momentum $n \in \mathbb{Z}$ and winding $w \in \mathbb{Z}$, and is spanned by vertex operators (see, e.g., Chapter 2 of \cite{Polchinski:1998rq})
\begin{equation}
    V_{n,w}(z,\bar{z}) = :\exp\left[ i\left( \left(\frac{n}{R} + w R \right) X(z) + \left(\frac{n}{R} - w R\right) \bar{X}(\bar{z}) \right) \right]:.
\end{equation}
The $U(1)^2$ symmetry acts on these states by
\begin{equation}
    V_{n,w} \mapsto e^{i\theta_m n} e^{i\theta_w w} V_{n,w},
\end{equation}
where $(\theta_m, \theta_w) \in U(1)^2$. This defines a representation of $U(1)^2$ on the state space, fibered over the moduli space $\mathbb{R}_{>0}$ parametrizing the radius $R$.

From the perspective of closed string states, the anomaly is seemingly not visible. Vertex operators transform under the $U(1)^2$ symmetry, and the corresponding symmetry structure depends only on additive charges---that is, on fusion rules---but not on any higher associativity data. As such, the closed string Hilbert space detects only the ring-level structure, leading to a principal $U(1)^2$-bundle over moduli space. The richer associator data---and hence the anomaly---seems invisible at this level.

In particular, we may consider the diagonal $\mathbb{Z}_2$ subgroup of $U(1)^2$ generated by simultaneously shifting both $\theta_m$ and $\theta_w$ by $\pi$. This corresponds to a $\mathbb{Z}_2$ symmetry transformation that simultaneously flips the signs of both the momentum and winding charges. Under this transformation,
\begin{equation}
    V_{n,w} \mapsto (-1)^{n+w} V_{n,w},
\end{equation}
which again probes only the fusion ring. Thus, the associated vector bundle structure is innocent about the anomaly, and no categorified structure over the moduli space emerges in the closed string sector.

\subsubsection{Stacks of D-brane (module) categories}
In contrast, the anomaly becomes manifest in the D-brane data. As discussed in Section~\ref{sect:anom:spacetime}, the mixed anomaly between $U(1)_m$ and $U(1)_w$ implies the emergence of a nontrivial 2-group gauge symmetry in the effective spacetime theory, with a modified Bianchi identity coupling the corresponding gauge fields. While the spacetime 2-group lives in the low-energy effective theory, we propose that its presence is reflected, at the worldsheet level, in the moduli space structure of the D-brane category.

We conjecture that the 2-group structure acts on the stack of D-brane categories over the moduli space.

To gain more control, we now focus on the specific diagonal $\mathbb{Z}_2$ subgroup of $U(1)^2$ generated by $\theta_m = \theta_w = \pi$. This transformation is a genuine subgroup of $U(1)^2$ that simultaneously shifts momentum and winding modes, and corresponds to a self-anomalous $\mathbb{Z}_2$ symmetry. The corresponding fusion category is $\mathrm{Vec}_{\mathbb{Z}_2}^\omega$, with nontrivial associator, schematically, $\omega(g,g,g) = -1$, with $g^2=1$ the nontrivial group element of $\mathbb{Z}_2$.

In this context, the D-brane category over moduli space truncates---insofar as symmetry structure is concerned---to a module category \cite{Ostrik:2001xnt} over $\mathrm{Vec}_{\mathbb{Z}_2}^\omega$. Importantly, this module category contains no fixed objects: Dirichlet branes $|D; x_0\rangle$ are invariant under $U(1)_w$ but transform under $U(1)_m$ as $x_0 \mapsto x_0 + \pi R$, while Neumann branes behave oppositely. No brane is invariant under both symmetries, nor under their diagonal $\mathbb{Z}_2$ subgroup.

Concretely, Dirichlet and Neumann boundary states are given by
\begin{equation}
    |D; x_0\rangle = \sum_{n \in \mathbb{Z}} e^{-i n x_0 / R} |n, w=0\rangle, \qquad |N; \tilde{x}_0\rangle = \sum_{w \in \mathbb{Z}} e^{-i w R \tilde{x}_0} |n=0, w\rangle,
\end{equation}
where $x_0$ and $\tilde{x}_0$ denote the position and Wilson line moduli of D0- and D1-branes, respectively. The diagonal $\mathbb{Z}_2$ symmetry maps $|D; x_0\rangle \mapsto |D; x_0 + \pi R\rangle$, and similarly acts nontrivially on Neumann branes via a shift of $\tilde{x}_0$.

Suppose, for contradiction, that the D-brane category were a module category over $\mathrm{Vec}_{\mathbb{Z}_2}$, i.e., that the symmetry were anomaly-free. Then, as in any anomaly-free setting, one would expect at least one brane invariant under $g$. The absence of such a brane implies that the symmetry acts freely and transitively on the set of simple branes---a hallmark of the \emph{regular} module category over $\mathrm{Vec}_{\mathbb{Z}_2}^\omega$. Physically speaking, the absence of $\mathbb{Z}_2$-symmetric brane boundary state is one of the defining property when the symmetry is anomalous\footnote{See, e.g.~\cite{Choi:2023xjw} for a detailed discussion on this physical perspective.}.

\begin{figure}
    \centering
    \includegraphics[width=10cm]{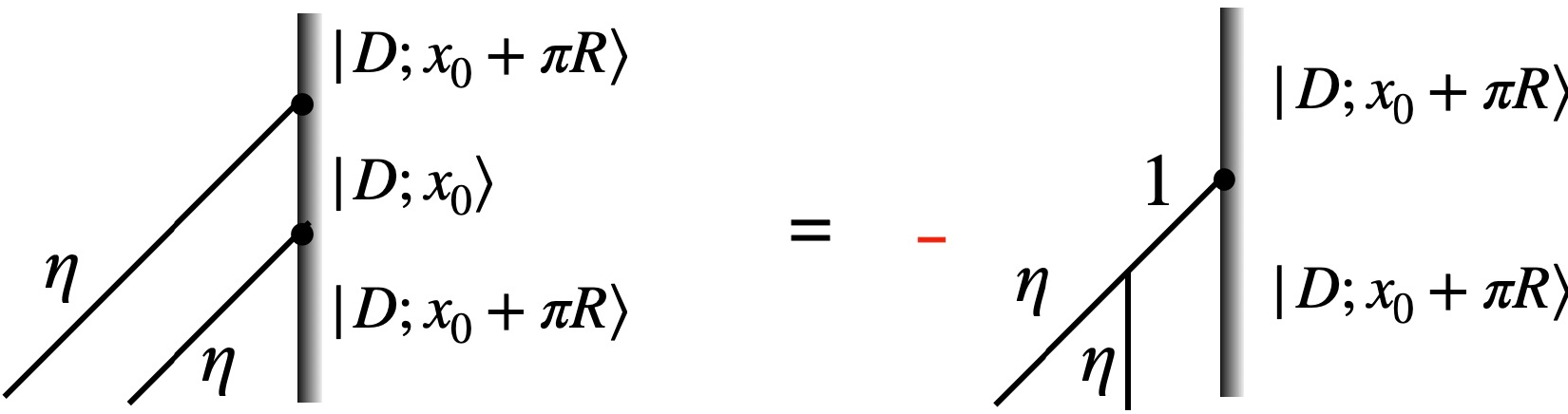}
    \caption{The associator of boundaries for D0-branes with the anomalous $\mathbb{Z}_2$ topological line $\eta$. The minus sign on the right-hand side represents the nontrivial associator phase, indicating the failure of strict associativity under the action of the $\mathbb{Z}_2$ line. This illustrates that with respect to the anomalous $\mathbb{Z}_2$ symmetry, i.e. $\mathrm{Vec}_{\mathbb{Z}_2}^\omega$ fusion category, the associated structure over moduli space for D-branes is a stack of module categories over $\mathrm{Vec}_{\mathbb{Z}_2}^\omega$.}
    \label{fig:twisted-z2-associator}
\end{figure}

This regular module category contains two simple objects $M_+$ and $M_-$, satisfying:
\begin{equation}
    \eta \otimes M_+ \cong M_-, \quad \eta \otimes M_- \cong M_+,
\end{equation}
where $\eta$ is the nontrivial simple object of $\mathrm{Vec}_{\mathbb{Z}_2}^\omega$, corresponding to the nontrivial group element $g\in \mathbb{Z}_2$.
These module category objects can be physically realized, for example, as 
\begin{equation}
\begin{split}
     &M_+: \text{D0-brane at}~x_0,\\
     &M_-: \text{D0-brane at}~x_0+\pi R.
\end{split}
\end{equation}
The key feature is that the associator of the module category over $\mathrm{Vec}_{\mathbb{Z}_2}^\omega$---the data specifying how iterated actions of $g$ associate---is nontrivial. In fact, since this is the regular module category, its associator \emph{coincides} with that of $\mathrm{Vec}_{\mathbb{Z}_2}^\omega$ itself. This associator is best understood pictorially: the nontrivial associativity data appears as a sign in a basic associative move, which we illustrate in Figure~\ref{fig:twisted-z2-associator}.

This associator---which can be explicitly represented diagrammatically (see Figure~\ref{fig:twisted-z2-associator})---encodes the failure of strict equivariance, and hence the anomaly, at the level of moduli space structure. The D-brane category must therefore be described as a stack of module categories over $\mathrm{Vec}_{\mathbb{Z}_2}^\omega$, glued together via this associator.

This example illustrates a general theme: closed string states define vector bundles over moduli space governed by fusion rings, which are insensitive to associators and thus to anomalies; in contrast, D-brane categories define stacks of module categories governed by fusion categories, in which associators---and hence anomalies---play a central role.

Before closing this section, let us quickly comment on structures over more involved CFT moduli space, specifically, over the Narain moduli space of flat connections for the symmetries.  To be a bit more general, suppose that instead of compactifying on a single $S^1$, we compactify on a $p$-dimensional torus $X = (S^1)^p$.  The spacetime theory contains a $G = U(1)^p \times U(1)^p$ gauge symmetry, generalizing the $U(1)_m \times U(1)_w$ above\footnote{One can also pursue a cohomological analysis similar to that of Eq.~(\ref{eq:adjcyc}), see \cite[Section 2]{Giotopoulos:2024jcr}.}.  Let ${\cal M}(X)$ denote the moduli space of flat $G$ connections on $X$.  It is a standard result \cite{tonypriv}
that over ${\cal M}(X) \times X$, there is a universal flat $G$-connection -- in essence,
a universal bundle of rank $2p$, reflecting the local spacetime and global worldsheet $G$ symmetry.
This is the prototype for the bundle which, for $G$ finite, would be described as a bundle of fusion rings of Vec$(G,\alpha)$.  The anomaly $\alpha$ is in principle encoded in a gerbe
structure, see e.g.~\cite{Bouwknegt:2011pq} for examples of gerbes over moduli spaces \footnote{Nonzero $H$ flux, as seen in the earlier construction, is known to be associated with nonassociativity on brane worldvolumes, see for example \cite{Cornalba:2001sm,Herbst:2001ai,Herbst:2003we}, \cite[section 2.5]{Szabo:2019hhg}, \cite{Szabo:2019gvu}, so in general it is extremely natural to link
$H$ flux and an associator $\alpha$.}. Understanding how such structures generalize to higher-categorical bundles and stacks for continuous or non-invertible symmetries remains an open and rich direction. 

In the remainder of this paper, we explore precisely such generalizations in the context of non-invertible symmetries arising in exceptional holonomy compactifications. The language developed here, especially the behavior under anomalous $\mathbb{Z}_2$ symmetries, will resurface in Section~\ref{sect:exchol} when we study non-invertible symmetries in exceptional holonomy SCFTs.

\section{Non-invertible symmetries}   \label{sect:noninv}

In the previous sections, we studied structures over moduli spaces induced by invertible global symmetries, both non-anomalous and anomalous. In this section, we turn to the more subtle case of non-invertible symmetries.

Motivated by our discussion in Section~\ref{sect:genl:anom} of the invertible fusion categorical symmetry $\mathrm{Vec}_{\mathbb{Z}_2}^\omega$ in the compact boson CFT, we propose that non-invertible worldsheet symmetries give rise to categorified structures over the SCFT moduli space, generalizing the bundles and stacks encountered in the invertible case. Concretely, for a fusion category $\mathcal{C}$:
\begin{itemize}
    \item The fusion ring of $\mathcal{C}$—obtained by forgetting its associator—acts via vector bundles over the moduli space of closed string states.
    \item The full fusion category $\mathcal{C}$ defines a stack\footnote{For more formal remarks on stacks and 2-vector bundles associated to categorical symmetries, see Appendix~\ref{app:stack-preliminaries}.} of categories over the moduli space, acting on stacks of D-brane categories via $\mathcal{C}$-module structure.
\end{itemize}

This general structure parallels the invertible symmetry story: the associator—or anomaly—data is invisible to the closed string state space, but governs the structure of the D-brane categories. It should reduce to the structures discussed previously when $\mathcal{C}$ contains an invertible subcategory (e.g.~$\mathrm{Vec}(G,\alpha)$ for finite $G$), and we expect it to provide a natural language for describing how global symmetry data organizes across families of (S)CFTs.

As in the case of anomalous invertible symmetries, one might hope that non-invertible symmetries on the worldsheet also admit a target-space avatar—perhaps in terms of categorified higher structures of gauge fields. However, such a spacetime interpretation remains poorly understood. A brief review of known constraints on gauging fusion category symmetries in two-dimensional QFTs is given in Appendix~\ref{app:gaugeability}, to help contextualize what can or cannot appear as spacetime structures.

That said, the situation may closely parallel the anomalous invertible case discussed in Section~\ref{sect:genl:anom}, where global worldsheet symmetries do not lift directly to ordinary spacetime gauge symmetries, but instead extend—e.g., to a 2-group—whose gauging governs the fiber structures over the SCFT moduli space. We expect a similar story may hold for non-invertible symmetries: even if no consistent spacetime gauge theory exists in the usual sense, the categorical symmetry visible on the worldsheet can still organize the fiber structure over moduli space. As in the anomalous setting, the structure over the moduli space remains well-defined and physically relevant, even if the corresponding spacetime picture is incomplete or not yet understood.

\section{Conjectures and evidence for $G_2$ and $\mathrm{Spin}(7)$ SCFTs}   \label{sect:exchol}

In this section, we illustrate the general proposal of Section~\ref{sect:noninv} through concrete examples: string compactifications on $G_2$ and $\mathrm{Spin}(7)$ manifolds, where non-invertible worldsheet symmetries of (tri)critical Ising type naturally arise. We explain how these symmetries lead to categorified structures over moduli space, generalizing the Bagger–Witten line bundle to cases without $U(1)_R$ symmetry.

\subsection{Non-invertible symmetries on $G_2$ and Spin$(7)$ SCFTs}

\subsubsection{Generalities}

Two-dimensional SCFTs describing manifolds of $G_2$ and Spin$(7)$ holonomy were discussed in \cite{Shatashvili:1994zw}.  In those theories, there is no global $U(1)_R$ symmetry, hence one does not expect a line bundle over the moduli space.  The largest invertible symmetry is the ${\mathbb Z}_2$ discrete $R$-symmetry of the two-dimensional $\mathcal{N}=1$ supersymmetry algebra\footnote{This $R$-symmetry acts by flipping the sign of the corresponding Majorana-Weyl fermions. See, e.g,  \cite{Sakamoto:1984zk, Hull:1985jv, Brooks:1986uh, Gukov:2019lzi, Franco:2021ixh} for more details of 2D minimally SUSY algebra and its representations.}.
However, the role played by the $U(1)_R$ current algebra in an $N=2$ SCFT is instead played by the tricritical Ising model (for $G_2$ manifolds) and
the ordinary Ising model (for Spin$(7)$ manifolds), both of which define non-invertible symmetries subsuming the ${\mathbb Z}_2$ discrete $R$-symmetry, and both of which contain an analogue of the spectral flow operator generating spacetime supersymmetry.  

Let us briefly review how the (tricritical) Ising sector shows up in ($G_2$) Spin$(7)$ SCFTs, following the original results in \cite{Shatashvili:1994zw}. Analogous to the Calabi-Yau case in Eq. (\ref{eq: u(1) from reduction}), where the principal $U(1)$ bundle can be regarded as a ``quotient'' of the holonomy group of K\"{a}hler by Calabi-Yau, one attempts to perform a similar quotient for general 8-dimensional (resp. 7-dimensional) real Riemannian target space with the holonomy group $SO(8)$ (resp. $SO(7)$) by that of Spin$(7)$ (resp. $G_2$). Although the result of the quotient does not make sense as a group, one can consider actions on the worldsheet fermions and substituting the quotient by the corresponding current algebra. The quotient then makes sense as a coset model:
\begin{equation}
 c\left( \frac{\text{Spin}(8)_1}{\text{Spin}(7)_1} \right)=\frac{1}{2}, ~ c\left( \frac{\text{Spin}(7)_1}{(G_2)_1} \right)=\frac{7}{10},
\end{equation}
Therefore, the role played by $U(1)$ in the Calabi-Yau case is now realized via rational chiral algebras with $c=\frac{1}{2}$ and $c=\frac{7}{10}$ for Spin$(7)$ and $G_2$, respectively. The corresponding RCFTs are uniquely Ising CFT ($c=\frac{1}{2}$) and tricritical Ising CFT ($c=\frac{7}{10}$) respectively. 

These RCFT sectors can be alternatively realized by introducing spectral flow operators to extend the conventional $\mathcal{N}=1$ superconformal algebra \cite{Shatashvili:1994zw}. In terms of the free-field realization where generators of the extended superconformal algebra can be written in terms of worldsheet bosons and fermions, spectral flow operators precisely correspond to the Cayley 4-form for Spin$(7)$, the associative 3-form and the coassociative 4-form for $G_2$, respectively. Geometrically, it is these calibrated forms responsible for the exceptional holonomy of Spin$(7)$ and $G_2$ compared to generic Riemannian manifolds. The RCFT sector is precisely generated by the spectral flow operator. For example, consider Spin$(7)$ SCFT with the spectral operator $I$ corresponding to the target space Cayley 4-form, one obtains the following OPE \cite{Shatashvili:1994zw}
\begin{equation}
   I(z)I(w)\sim \frac{16}{(z-w)^4}+\frac{16}{(z-w)^2}I(w)+\frac{8}{(z-w)}\partial I(w).
\end{equation}
Therefore, one can construct $T_I \equiv \frac{1}{8}I(z)$ and realize exactly the $c=\frac{1}{2}$ Virasoro algebra for the Ising CFT. Similar procedure works for $G_2$ SCFTs.

The non-invertible symmetries can then be realized by Verlinde lines associated to the Ising and the tricritical Ising chiral algebras for the Spin$(7)$ and $G_2$ SCFTs, respectively. For Ising sector, there are three Verline lines $1, \eta$ and $\mathcal{D}$ in the Ising category, with the following fusion rules
\begin{equation}\label{eq:fusion rule of Ising}
 \mathcal{D}\times \mathcal{D}=1+\eta, ~~\eta\times \eta=1, ~\eta\times \mathcal{D}=\mathcal{D}\times \eta=\mathcal{D}.
\end{equation}
Mathematically, this is identified as the TY$(\mathbb{Z}_2)$ Tambara-Yamagami fusion ring \cite{TAMBARA1998692}. 
For $G_2$, there are six Verlinde lines $1, \eta, \mathcal{D}, W, \eta W, W \mathcal{D}$ in the tricritial Ising category, with the following fusion rules
\begin{equation}\label{eq: fusion rule of tricritical Ising}
\begin{split}
        &\mathcal{D}\times \mathcal{D}=1+\eta, ~~\eta\times \eta=1, ~\eta\times \mathcal{D}=\mathcal{D}\times \eta=\mathcal{D},\\
    &W\times W=1+W.
\end{split}
\end{equation}
The first line is again the TY$(\mathbb{Z}_2)$ fusion ring as in the Ising category, while the second line is the Fibonacci fusion ring.

To construct topological line operators generating these symmetries for the full Spin$(7)$ and $G_2$ SCFTs, the key observation is to notice that the full Virasoro algebra decomposes into a RCFT sector and the rest sector, commuting with RCFT sector. For example, in Spin$(7)$ SCFTs, labeling the Ising sector as $T_I$ and the rest sector as $T_r$, one observes \cite{Shatashvili:1994zw}:
\begin{equation}
    T=T_I+T_r, ~~T_I(z)T_r(w)\sim 0.
\end{equation}
Therefore, one can build topological operators 
\begin{equation}
    L_I\otimes 1_r
\end{equation}
where $L_I=1, \eta$ and $\mathcal{D}$ are Verlinde lines for the Ising sector, while $1$ is the identity line operator for the $T_r$ sector. This composed line operator is obviously satisfying the topological condition
\begin{equation}
    (L_I\otimes 1_r) T =T (L_I\otimes 1_r),
\end{equation}
thus generating a symmetry for the Spin$(7)$ SCFT. Similar argument works for $G_2$. 

For simplicity, in the following discussion we will omit ``$\otimes 1_r$'' and just use the Verlinde line itself to label the topological line of the Spin$(7)$ and $G_2$ symmetries. The actions of these non-invertible symmetries on Spin$(7)$ and $G_2$ SCFTs, including closed string states and D-brane boundary states, will be investigated in Section~\ref{sect:cohom} and Section~\ref{subsec: exceptional d-brane categories}.

\subsubsection{Basic conjectures}
By analogy with invertible symmetries, 
one then expects that over a moduli space of SCFTs describing a family of $G_2$ or Spin$(7)$ manifolds, there will be a stack of fusion categories, reflecting that symmetry, and a rigidification, a bundle of fusion rings, of which the analogue of the spectral flow operator is a section of an associated vector bundle.  Because of the relation to spectral flow, the stack over the SCFT moduli space should act as analogues of the Bagger-Witten and Hodge line bundles on a moduli space of Calabi-Yau varieties. Furthermore, the full fusion category defines a stack over moduli space, acting on stacks of D-brane categories via module category structure.
However, the details are a bit more subtle than for Bagger-Witten line bundles over moduli spaces of two-dimensional (2,2) supersymmetric SCFTs.

We discussed several subtleties in understand the this phenomenon in the target-space theory in the previous section.  One of those subtleties is particularly acute in
Spin$(7)$ compactifications.  There,
the low-energy theory is two-dimensional, and we believe we understand the basics of gauging in two-dimensional theories.
However, the fusion category symmetry arising in a Spin(7) SCFT is the Ising category, which is not gaugeable, as observed in e.g.~\cite[section 2.4]{Perez-Lona:2023djo}.  
However, as discussed in the example of anomalous worldsheet $\mathbb{Z}_2$ symmetry of the $S^1$ compactification of the bosonic string whose target space gauge symmetry is a 2-group, although there can not be gauged Ising categorical symmetry in compactified two-dimensional spacetime theory, we conjecture there still could be a higher-categorical extension so that Ising categorical symmetry is part of a larger gauge symmetry structure (at least at string tree level). The detailed investigation of this higher-categorical spacetime realization is out of the scope of this paper, and we will focus on implications of worldsheet symmetries.

Another subtlety in this proposal involves the number of Bagger-Witten bundles.
As reviewed in section~\ref{sect:genl:nonanom}, over moduli spaces of Calabi-Yau manifolds, the Bagger-Witten line bundles can be interpreted as bundles of covariantly constant spinors, and that because a Calabi-Yau threefold has two different covariantly constant spinors, there are two different Bagger-Witten line bundles.

On $G_2$ and Spin$(7)$ manifolds, one could imagine a similar geometric definition.
In these cases, there is only a single covariantly constant spinor \cite{bryantpriv}, reflecting the single operator in the (tricritical) Ising model which generates spacetime supersymmetry.  Since there is only a single covariantly constant spinor, we expect only a single analogue of the Bagger-Witten line bundle.  Another consequence of having only a single covariantly constant spinor is that the physical argument above for flatness from \cite{Donagi:2019jic} need not apply to the fusion category 'bundles' (stacks) we propose here.  That said, fusion categories are closely analogous to finite groups, hence presumably
fusion category 'bundles' should be closely analogous to bundles whose structure groups are finite groups,
which are automatically flat.

In Section~\ref{sect:cohom}, we will give this proposal some teeth, by computing the action of those non-invertible symmetries on the cohomology of Spin(7) and $G_2$ holonomy spaces. This serves as a geometric counterpart of the closed string vector space, fibered over the moduli space, acted on by non-invertible symmetries. In Section~\ref{subsec: exceptional d-brane categories}, we characterize D-branes via module categories over fusion categories, implying the stack structure associated to D-brane categories acted upon by non-invertible symmetries.  These two perspectives provide better understanding of the global structure being proposed.

\subsection{Action on closed string state space as cohomology}   \label{sect:cohom}

So far we have argued abstractly that moduli spaces of SCFTs describing $G_2$ and Spin$(7)$ manifolds should carry a stack of fusion categories,
with transition functions defined by the fusion algebra of non-invertible symmetries of the tricritical Ising model and the ordinary Ising model, respectively, essentially because the Verlinde lines of the Ising and the tricritical Ising chiral algebras for the Spin$(7)$ and
$G_2$ theories define topological line operators.

In this section, we shall describe the actions of those categories (more precisely, their fusion rings) on the cohomology of the spaces, to help understand more concretely the role of the proposed vector bundles of closed string state space and their signatures. We will see that the cohomology forms a sum of one-dimensional representations of the fusion category in each case.  (We will also see that the action on the cohomology does not preserve product structures, and so does not act by ring endomorphisms.)

As the analysis of the Spin(7) case is easier than that of $G_2$, we begin with
Spin(7).

\subsubsection{Spin$(7)$}

In the case of SCFTs describing Spin$(7)$ manifolds, the role of the $U(1)_R$ current is replaced by
the CFT for the Ising model \cite{Shatashvili:1994zw}.

Following \cite[section 5]{Ginsparg:1988ui}, the ordinary Ising model is a $c=1/2$ bosonic minimal model with Virasoro primary fields corresponding to the identity,
$\sigma$ of conformal weights $(1/16, 1/16)$ corresponding to spin,
and $\epsilon$ of conformal weights $(1/2, 1/2)$ corresponding to energy.
These local operators obey the fusion algebra \cite[equ'n (5.4)]{Ginsparg:1988ui}
\begin{equation}    \label{eq:ising:fusion}
    \sigma \times \sigma = 1 + \epsilon, \: \: \:
    \sigma \times \epsilon = \sigma, \: \: \:
    \epsilon \times \epsilon = 1.
\end{equation}
The simple lines of this theory are in one-to-one correspondence \cite{Petkova:2000ip} with the Virasoro primaries,
hence there are three lines, including the identity.

Note that the fusion rules above include a single ${\mathbb Z}_2$, generated by $\epsilon$.
As this is invertible, it should define a ${\mathbb Z}_2$ structure over the moduli space.
(It also survives at nonzero string coupling.)
Our conjecture, as applied to Spin$(7)$, extends that ${\mathbb Z}_2$ structure to the larger Ising structure containing it.

In \cite[section 3.2]{Shatashvili:1994zw}, 
it is noted that the energy operator $\epsilon$ acts
as a precise analogue of the $N=2$ spectral flow operator, in the sense that
it exchanges the Ramond ground state and a certain NS highest weight state.
Specifically, the chiral R sector states are given in \cite[equ'n (3.44)]{Shatashvili:1994zw} as
\begin{equation}
    R:  \: \: \:
    | 1/2, 0 \rangle, \: \: \:
    | 0, 1/2 \rangle, \: \: \:
    | 1/16, 7/16 \rangle,
\end{equation}
and the corresponding NS states are \cite[equ'n (3.45)]{Shatashvili:1994zw}
\begin{equation}
    NS: \: \: \:
    | 0, 0 \rangle, \: \: \:
    | 1/2, 1/2 \rangle, \: \: \:
    | 1/16, 7/16 \rangle,
\end{equation}
both labelled by $| \Delta_I, \Delta_r \rangle$ where $\Delta_I$ is the conformal weight under the Ising piece $T_I$ of the stress tensor $T = T_I + T_r$, and $\Delta_r$ the weight under the remainder $T_r$.  For example, using the fusion rules~(\ref{eq:ising:fusion}), one quickly finds that $\epsilon$ acts on states as
\begin{eqnarray*}
    | 1/2, 0 \rangle & \mapsto & | 0, 0 \rangle, 
    \\
    | 0, 1/2 \rangle & \mapsto & |1/2, 1/2 \rangle,
    \\
    | 1/16, 7/16 \rangle & \mapsto & | 1/16, 7/16 \rangle,
\end{eqnarray*}
using $\epsilon \times \epsilon = 1$, $\epsilon \times 1 = \epsilon$, and
$\epsilon \times \sigma = \sigma$, respectively.

Furthermore, the Ising stress tensor $T_I$ piece of the stress tensor is proportional to the spin-2 operator $\hat{X}$ corresponding to the Cayley 4-form. 
Also, from \cite{Shatashvili:1994zw}, $(-)^F$ is realized by the ${\mathbb Z}_2$ Ising symmetry $\sigma \mapsto - \sigma$ (leaving $1$ and $\epsilon$ invariant).

In addition, we observe that the Ising lines define topological symmetries of the entire theory.  Given an Ising line $L$, it acts on the state space ${\cal H}_I \otimes {\cal H}_r$ as, $L \otimes 1$.

Next, we will compute the action of the Verlinde line operators (corresponding to the Ising operators) on the cohomology classes of a Spin(7) manifold, to find a mathematical consequence of the proposed fusion algebra bundle.

A simple Verlinde line operator $L_i$ acts on a Virasoro primary state $| \phi_j \rangle$ (and its descendants) as
\cite[section 4.2.1]{Choi:2023xjw}
\begin{equation}  \label{eq:lines-prims}
    L_i | \phi_j \rangle \: = \: \frac{ S_{ij} }{ S_{0j} } | \phi_j \rangle,
\end{equation}
where $S$ denotes the modular $S$ matrix.  It is easy to check this
is consistent with the fusion rules
\begin{equation}
    L_i \otimes L_j \: = \: \sum_k N^k_{ij} \, L_k
\end{equation}
using the diagonalization
\begin{equation}
    N_{ij}^k \: = \: \sum_m \frac{ S_{im} S_{jm} S_{mk}^* }{ S_{0m} }.
\end{equation}

For purposes of computing the actions of the lines on the Virasoro primaries,
the modular $S$ matrix, acting on the chiral primaries in the order $(1,\epsilon,\sigma)$,
is \cite{Moore:1988qv} (see also \cite[equ'n (10.138)]{DiFrancesco:1997nk})
\begin{equation}
    \frac{1}{2} \left[ \begin{array}{ccc} 
    1 & 1 & + \sqrt{2} \\
    1 & 1 & - \sqrt{2} \\
    +\sqrt{2} & - \sqrt{2} & 0 \end{array} \right]
\end{equation}
From equation~(\ref{eq:lines-prims}) one can compute the action of the Verlinde lines on the parimaries, which we list in the table below:
\begin{center}
    \begin{tabular}{c|ccc}
    & $1$ & $|\epsilon\rangle$ & $|\sigma\rangle$ \\ \hline
    $1$ & $1$ & $1$ & $1$ \\
    $L_{\epsilon}\equiv \eta$ & $1$ & $1$ & $-1$ \\
    $L_{\sigma}\equiv \mathcal{D}$ & $\sqrt{2}$ & $- \sqrt{2}$ & $0$
    \end{tabular}
\end{center}
For example, $\eta | \sigma \rangle = - | \sigma \rangle$.

We interpret an Ising line $L$ acting on a chiral state in ${\cal H}_I \otimes {\cal H}_r$ to be, strictly speaking, $L \otimes 1$, where $L$ acts on ${\cal H}_I$ and
$1$ acts on ${\cal H}_r$.  (Note that this means the Ising lines automatically define topological lines.)
The action of lines on chiral R sector states can then be read off as
\begin{eqnarray}
    \eta: & | 1/2, 0 \rangle & \mapsto \: | 1/2, 0 \rangle,  
    \label{eq:r:Lepsilon:1}
    \\
    & | 0, 1/2 \rangle & \mapsto \: | 0, 1/2 \rangle,
    \\
    & | 1/16, 7/16 \rangle & \mapsto \: - | 1/16, 7/16 \rangle,
    \\
    \mathcal{D}: & | 1/2, 0 \rangle & \mapsto \: - \sqrt{2}\, | 1/2, 0 \rangle,
    \\
    & | 0, 1/2 \rangle & \mapsto \: + \sqrt{2} \,| 0, 1/2 \rangle,
    \\
    & | 1/16, 7/16 \rangle & \mapsto \: 0,
    \label{eq:r:Lsigma:sigma}
\end{eqnarray}
and the action of lines on chiral NS sector states is similarly
\begin{eqnarray}
    \eta: & |0, 0\rangle & \mapsto \: | 0, 0\rangle,
    \\
    & | 1/2, 1/2 \rangle & \mapsto \: | 1/2, 1/2 \rangle, 
    \\
    & | 1/16, 7/16 \rangle & \mapsto \: - | 1/16, 7/16 \rangle,
    \\
    \mathcal{D}: & | 0, 0 \rangle & \mapsto \: \sqrt{2}\, | 0, 0 \rangle,
    \\
    & | 1/2, 1/2 \rangle & \mapsto \:  -\sqrt{2} \, |1/2, 1/2 \rangle,
    \\
    & | 1/16, 7/16 \rangle & \mapsto \: 0.
\end{eqnarray}

Next, given the actions of the lines on the chiral states, we can compute the action of lines on cohomology groups, and argue that the cohomology of a Spin(7) manifold naturally transforms in a representation of the Ising fusion algebra, specifically, a sum of products of one-dimensional
representations.

To that end, let us interpret the actions on chiral states above in terms of the cohomology of the Spin(7) manifold.  Following \cite[section 3]{Shatashvili:1994zw}, the nonchiral RR states are related to cohomology as in the table below:
\begin{center}
    \begin{tabular}{c|c}
    States & Cohomology \\ \hline
    $ | 1/2, 0 \rangle_L \otimes | 1/2, 0 \rangle_R$ & $H^0 = H^8$; unique up to rescaling
    \\
    $ | 0, 1/2 \rangle_L \otimes | 1/16, 7/16 \rangle_R$,
    $ |1/16, 7/16 \rangle_L \otimes |0, 1/2 \rangle_R$ & $H^3 = H^5$ 
    \\
    $| 0, 1/2 \rangle_L \otimes |0, 1/2\rangle_R$,
    $| 1/16, 7/16 \rangle_L \otimes | 1/16, 7/16 \rangle_R$ & $H^2 \oplus H^{4 -} = H^6 \oplus H^{4 +}$
    \end{tabular}
\end{center}
(As we have only captured quantum numbers, for example the states we are labelling
$|0, 1/2\rangle_L \otimes | 1/16, 7/16 \rangle_R$ can appear with multiplicity.)
For example, the ${\mathbb Z}_2$ symmetry generated by $\sigma \mapsto - \sigma$, $1, \epsilon$ invariant,
measures whether the cohomological degree is even or odd, which is reflected above.

Following \cite[section 3]{Shatashvili:1994zw}, we assume that the manifold is connected ($b_0 = 0)$, and that it does not have enhanced supersymmetry (hence $b_1 = 0 = b_7$).

Then, from equations~(\ref{eq:r:Lepsilon:1})-(\ref{eq:r:Lsigma:sigma}), we find that
under the action of the lines $\eta$, $\mathcal{D}$, the cohomology transforms as follows:
\begin{center}
    \begin{tabular}{c|c|c}
    Cohomology & $\eta$ eigenvalue & $\mathcal{D}$ eigenvalue \\ \hline
    $H^0 = H^8$ & $+1$ & $+2$ \\
    $H^3 = H^5$ & $-1$ & $0$ \\
    $H^2 \oplus H^{4-} = H^6 \oplus H^{4+}$ & $+1$ & (*)
    \end{tabular}
\end{center}
The entry $(*)$ indicates an ambiguity: $| 0, 1/2 \rangle_L \otimes | 0, 1/2 \rangle_R$ has $\mathcal{D}$ eigenvalue $+2$, whereas $ | 1/16, 7/16 \rangle_L \otimes |1/16, 7/16 \rangle_R$ has $\mathcal{D}$ eigenvalue $0$.
As states of these quantum numbers span $H^2 \oplus H^{4-}$, one reasonable interpretation
is that the $\mathcal{D}$ eigenvalue distinguishes $H^2$ from $H^{4-}$, so that one cohomology group transforms with eigenvalue $+2$ and the other with eigenvalue $0$.  
In passing, we note that $\eta$ is the ${\mathbb Z}_2$ described earlier.

Note that the Cayley four-form is a self-dual four-form, hence lies amongst the states
classified by $H^6 \oplus H^{4+}$.

We observe that states with $\eta$ eigenvalue $+1$ and $\mathcal{D}$ eigenvalue $2$ should be interpreted as invariant under the symmetry.
As noted in \cite[section 5.1]{Perez-Lona:2023djo}, to be `invariant' under a non-invertible symmetry means that lines multiply operators by quantum dimensions.  Since $\epsilon$ has quantum dimension $1$ and $\sigma$ has quantum dimension $\sqrt{2}$, and states are formed from the tensor product of both left- and right-moving sectors, we infer that
an invariant must have eigenvalue $(\sqrt{2})^2 = 2$ under $L_{\sigma} = \mathcal{D}$.

In passing, from the $\mathcal{D}$ eigenvalues, it should be clear that the action does not preserve products, and so is not a ring endomorphism.

As a consistency check, let us compute one-dimensional representations of the 
fusion algebra of $\{1, \epsilon, \sigma\}$, to check that the cohomology transforms under such representations.  To that end, let $| \phi \rangle$ be a state generating a one-dimensional representation, and write
\begin{equation}
    \eta | \phi \rangle \: = \: \alpha | \phi \rangle, \: \: \:
    \mathcal{D} | \phi \rangle \: = \: \beta | \phi \rangle.
\end{equation}
Then, from the fusion rules
\begin{equation}
    \eta \times \eta = 1, \: \: \:
    \mathcal{D} \times \eta = \mathcal{D}, \: \: \:
    \mathcal{D} \times \mathcal{D} = 1 + \eta,
\end{equation}
we get the corresponding algebraic constraints on $\alpha$ and $\beta$:
\begin{equation}
    \alpha^2 = +1, \: \: \:
    \alpha \beta = \beta, \: \: \:
    \beta^2 = 1 + \alpha,
\end{equation}
respectively.
We find three solutions:
\begin{equation}
(\alpha = -1, \beta = 0), \: \: \:
(\alpha = +1, \beta = \pm \sqrt{2}).
\end{equation}
Now, in principle, the cohomology groups should transform in a product of two (one-dimensional) representations,
one for each chiral factor, and indeed, it is trivial to check that they do, as expected.

\subsubsection{$G_2$}

First, let us recall the possible non-invertible symmetries of the tricritical Ising model,
as that will play the same role in $G_2$ holonomy manifolds as the $U(1)_R$ symmetry does for
ordinary K\"ahler targets.

From \cite{Friedan:1983xq,Friedan:1984rv,Friedan:1986kd},
\cite[equ'n (24)]{Saura-Bastida:2024yye},
the tricritical Ising model is a RCFT of $c = 7/10$, containing 
\cite[section 4.2.1]{Choi:2023xjw}
six primary
operators $1, \epsilon, \epsilon', \epsilon'', \sigma, \sigma'$, and six lines (in one-to-one correspondence \cite{Petkova:2000ip}).

The Virasoro primaries have dimension \cite[table 7.2]{DiFrancesco:1997nk}
\begin{center}
\begin{tabular}{c|cccccc}
Primary & $1$ & $\epsilon$ & $\epsilon'$ & $\epsilon''$ & $\sigma$ & $\sigma'$ \\ \hline
Dimension & $0$ & $1/10$ & $3/5$ & $3/2$ & $3/80$ & $7/16$
\end{tabular}
\end{center}
and nontrivial fusion rules \cite[table 7.3]{DiFrancesco:1997nk}
\begin{equation}  \label{eq:tricrit:f1}
    \epsilon \times \epsilon = 1 + \epsilon', \: \: \:
    \epsilon \times \epsilon' = \epsilon + \epsilon'', \: \: \:
    \epsilon \times \epsilon'' = \epsilon', \: \: \:
    \epsilon' \times \epsilon' = 1 + \epsilon', \: \: \:
    \epsilon' \times \epsilon'' = \epsilon,
    \: \: \:
    \epsilon''\times \epsilon''=1,  
    %\: \: \:
\end{equation}
\begin{equation}  \label{eq:tricrit:f2}
    \epsilon \times \sigma = \sigma + \sigma', \: \: \:
    \epsilon \times \sigma' = \sigma, \: \: \:
    \epsilon' \times \sigma = \sigma + \sigma', \: \: \:
    \epsilon' \times \sigma' = \sigma, \: \: \:
    \epsilon'' \times \sigma = \sigma, \: \: \:
    \epsilon'' \times \sigma' = \sigma', 
\end{equation}
\begin{equation}  \label{eq:tricrit:f3}
    \sigma \times \sigma = 1 + \epsilon + \epsilon' + \epsilon'', \: \: \:
    \sigma \times \sigma' = \epsilon + \epsilon', \: \: \:
    \sigma' \times \sigma' = 1 + \epsilon''.
\end{equation}

Gathered as scaling fields, we have in the NS sector the superpairs $(1,\epsilon'')$, and $(\epsilon',\epsilon)$, as stated in \cite[p. 13]{Shatashvili:1994zw}, with assignments $[1]^+, [\epsilon'']^-=[\tfrac{3}{2}]^- , [\epsilon']^= = [\tfrac{3}{5}]^+,[\epsilon]^- = [\tfrac{1}{10}]^-$ for the operator $(-)^{F_I}$ in the tricritical Ising model. This is relevant as it is argued that in the $G_2$ theory $(-)^F = (-)^{F_I}$.  

Note that the fusion rules above include a single ${\mathbb Z}_2$, generated by $\epsilon''$.
As this is invertible, it should define a ${\mathbb Z}_2$ structure over the moduli space.
(It also survives at nonzero string coupling.)
Our conjecture, as applied to $G_2$, extends that ${\mathbb Z}_2$ structure to the larger tricritical Ising structure containing it.

As explained in \cite{Shatashvili:1994zw}, the spectral flow operator here is $\sigma'$, of conformal dimension $7/16$.

The lines are the identity, an invertible ${\mathbb Z}_2$ line
$\eta = L_{\epsilon''}$, 
and four other simple lines $W = L_{\epsilon'}$, $\eta W = L_{\epsilon}$, ${\cal D} = L_{\sigma'}$, $W {\cal D} = L_{\sigma}$,
with fusion rules
\begin{equation}
\eta \times \eta = 1, \: \: \: 
{\cal D} \times {\cal D} = 1 + \eta, \: \: \:
\eta \times {\cal D} = {\cal D} \times \eta = {\cal D}, \: \: \:
W \times W = 1 + W.
\end{equation}
The lines $\eta$, ${\cal D}$, $W$ have quantum dimensions
$1$, $\sqrt{2}$, $(1 + \sqrt{5})/2$, respectively.
Note this contains a Fibonacci subcategory $\{1, W\}$ and a 
${\mathbb Z}_2$ Tambara-Yamagami
subcategory $\{1, \eta, {\cal D} \}$ (the Ising category of the previous section).

Also, just as in the previous section, the tricritical Ising lines define topological symmetries of the entire theory.  Given a tricritical Ising line $L$, it acts on the state space ${\cal H}_I \otimes {\cal H}_r$ as, $L \otimes 1$.

A simple line operator $L_i$ acts on a Virasoro primary state $| \phi_j \rangle$ as
\cite[section 4.2.1]{Choi:2023xjw}
\begin{equation}
    L_i | \phi_j \rangle \: = \: \frac{ S_{ij} }{ S_{0j} } | \phi_j \rangle,
\end{equation}
where $S$ denotes the modular $S$ matrix.

For purposes of computing the actions of the lines on the Virasoro primaries,
the modular $S$ matrix, acting on the chiral primaries in the order $(1, \epsilon, \epsilon', \epsilon'', \sigma, \sigma')$,
is \cite[equ'n (10.139)]{DiFrancesco:1997nk}
\begin{equation}
    \frac{1}{\sqrt{5}} \left[ \begin{array}{cccccc}
    s_2 & + s_1 & + s_1 & s_2 & + \sqrt{2} s_1 & + \sqrt{2} s_2 \\
    s_1 & - s_2 & - s_2 & s_1 & + \sqrt{2} s_2 & - \sqrt{2} s_1 \\
    s_1 & - s_2 & - s_2 & s_1 & - \sqrt{2} s_2 & + \sqrt{2} s_1 \\
    s_2 & + s_1 & + s_1 & s_2 & - \sqrt{2} s_1 & - \sqrt{2} s_2 \\
    \sqrt{2} s_1 & + \sqrt{2} s_2 & - \sqrt{2} s_2 & - \sqrt{2} s_1 & 0 & 0 \\
    \sqrt{2} s_2 & - \sqrt{2} s_1 & + \sqrt{2} s_1 & - \sqrt{2} s_2 & 0 & 0
    \end{array}
    \right]
\end{equation}
where
\begin{equation}
    s_1 \: = \: \sin(2 \pi / 5), \: \: \:
    s_2 \: = \: \sin( 4 \pi / 5).
\end{equation}

This gives the action of the Verlinde lines on the primaries (also given in \cite[Table I]{Saura-Bastida:2024yye}):
\begin{center}
    \begin{tabular}{c|c|c|c|c|c|c}
 & 1 & $\epsilon$ & $\epsilon'$ & $\epsilon''$ & $\sigma$ & $\sigma'$
 \\ \hline
 1 & 1 & 1 & 1 & 1 & 1 & 1
 \\
 $\eta W = L_{\epsilon}$ & $\varphi$ & $-\varphi^{-1}$ & $-\varphi^{-1}$ & $\varphi$ & $\varphi^{-1}$ & $-\varphi$
 \\
 $W = L_{\epsilon'}$ & $\varphi$ & $-\varphi^{-1}$ & $-\varphi^{-1}$ & $\varphi$ & $-\varphi^{-1}$ & $\varphi$
 \\
 $\eta = L_{\epsilon''}$ & $1$ & $1$ & $1$ & $1$ & $-1$ & $-1$
 \\
 $W{\cal D} = L_{\sigma}$ & $\sqrt{2}\varphi$ & $\sqrt{2}\varphi^{-1}$ & $-\sqrt{2}\varphi^{-1}$ & $-\sqrt{2}\varphi$ & $0$ & $0$
 \\
 ${\cal D} = L_{\sigma'}$ & $\sqrt{2}$ & $-\sqrt{2}$ & $\sqrt{2}$ & $-\sqrt{2}$ & $0$ & $0$
\end{tabular}
\end{center}
for $\varphi= s_1 / s_2 = (1+\sqrt{5})/2$ the golden ratio (i.e.~the quantum dimension $\text{dim}(W)$).

We can use this to read off the actions on the ground states in the $G_2$ theory \cite[Eq. (3.14), (3.17)]{Shatashvili:1994zw}
\begin{eqnarray}
    \text{R}&:& \left\vert \left. \frac{7}{16}, 0 \right. \right\rangle = \vert \sigma',0\rangle; 
    \: 
    \left\vert \left. \frac{3}{80}, \frac{2}{5}\right. \right\rangle = \left\vert \left. \sigma, \frac{2}{5} \right. \right\rangle,
    \\
    \text{NS}&:& \vert 0, 0\rangle = \vert 1,0\rangle; 
    \:
    \left\vert \left. \frac{1}{10}, \frac{2}{5} \right. \right\rangle = \left\vert \left. \epsilon, \frac{2}{5} \right. \right\rangle,
\end{eqnarray}
for $|\Delta_I,\Delta_R\rangle$ with $\Delta_I$ the tricritical weight, and $\Delta_R$ the weight for the remaining part of the $G_2$ algebra.  The eigenvalues of the line operators on the R states above are listed in the table below:
\begin{center}
    \begin{tabular}{c|c|c}
    & $| 7/16, 0 \rangle$ & $ | 3/80, 2/5 \rangle$ \\ \hline
    $1$ & $1$ & $1$ \\
    $\eta W$ & $-\varphi$ & $\varphi^{-1}$ \\
    $W$ & $\varphi$ & $- \varphi^{-1}$ \\
    $\eta$ & $-1$ & $-1$ \\
    $W\mathcal{D}$ & $0$ & $0$ \\
    $\mathcal{D}$ & $0$ & $0$
    \end{tabular}
\end{center}

The (nonchiral) RR states are listed in \cite[equ'n (3.20)]{Shatashvili:1994zw} as
\begin{equation}
    | (7/16, 0)_L; (7/16, 0)_R; \pm \rangle, \: \: \:
    | (3/80, 2/5)_L; (3/80, 2/5)_R; \pm \rangle.
\end{equation}
The $\pm$ indicates a pair of choices in the non-Ising part of the conformal field theory,
with different $(-)^F$ eigenvalues, so that the $+$ states correspond to even-degree cohomology and the $-$ states correspond to odd-degree cohomology.

Following \cite[section 3]{Shatashvili:1994zw}, we assume that the manifold is connected ($b_0 = 0)$, and that it does not have enhanced supersymmetry (hence $b_1 = 0 = b_6$).
We therefore take the $| (7/16, 0)_L; (7/16, 0)_R; + \rangle$ state to be unique.
Putting this together, and following  \cite[section 3]{Shatashvili:1994zw},
we get the following dictionary between states and cohomology:
\begin{center}
    \begin{tabular}{c|c}
    States & Cohomology \\ \hline
    $| (7/16, 0)_L; (7/16, 0)_R; + \rangle$ & $H^0$; unique up to rescaling \\
    $| (3/80, 2/5)_L; (3/80, 2/5)_R; + \rangle$ & $H^2 \oplus H^4$ \\
    $| (3/80, 2/5)_L; (3/80, 2/5)_R; - \rangle$ & $H^3 \oplus H^5$ \\
    $| (7/16, 0)_L; (7/16, 0)_R; - \rangle$ & $H^7$; unique up to rescaling
    \end{tabular}
\end{center}
(As we have only captured quantum numbers, states above can appear with multiplicity, unless otherwise indicated.)

Given the dictionary above and our earlier results on line operator actions on chiral R sector states, we can read off the action of the lines on the cohomology.
Listed below, for each set of cohomology classes, are the eigenvalues under the
actions of various line operators:
\begin{center}
    \begin{tabular}{c|c|c|c|c|c}
    Cohomology & $\eta W$ & $W$ & $\eta$ & $W\mathcal{D}$ & $\mathcal{D}$ \\ \hline
    $H^0$ & $\varphi^2$ & $\varphi^2$ & $1$ & $0$ & $0$ \\
    $H^2 \oplus H^4$ & $\varphi^{-2}$ & $\varphi^{-2}$ & $1$ & $0$ & $0$ \\
    $H^3 \oplus H^5$ & $\varphi^{-2}$ & $\varphi^{-2}$ & $1$ & $0$ & $0$ \\
    $H^7$ & $\varphi^2$ & $\varphi^2$ & $1$ & $0$ & $0$ 
    \end{tabular}
\end{center}
where as before $\varphi = (1 + \sqrt{5})/2$.

Next, as a consistency check, let us compare to one-dimensional representations of the tricritical Ising model.  Define the eigenvalues $\alpha$, $\alpha'$, $\alpha''$, $\beta$, $\beta'$ by
\begin{equation}
    \eta W | \phi \rangle = \alpha | \phi \rangle, \: \: \:
    W | \phi \rangle = \alpha' | \phi \rangle, \: \: \:
    \eta | \phi \rangle = \alpha'' | \phi \rangle, \: \: \:
    W\mathcal{D} | \phi \rangle = \beta | \phi \rangle, \: \: \:
    \mathcal{D} | \phi \rangle = \beta' | \phi \rangle,
\end{equation}
and applying the fusion rules~(\ref{eq:tricrit:f1})-(\ref{eq:tricrit:f3}),
we find the following two sets of solutions:
\begin{enumerate}
    \item $\alpha = (1/2) (-1 \pm \sqrt{5}) = - \varphi, +\varphi^{-1}$, $\alpha' = - \alpha$, $\alpha'' = -1$, $\beta = 0 = \beta'$,
    \item $\alpha = (1/2)(+1 \pm \sqrt{5})$, $\alpha' = + \alpha$, $\alpha'' = +1$, 
    $\beta = \alpha \beta'$, $\beta' = \pm \sqrt{2}$.
\end{enumerate}
(In the second option, the choice of sign in $\beta'$ is independent of the choice of sign in $\alpha$, so the second option describes four choices.)  As expected, all of the cohomology is in the square of a one-dimensional representation, namely the square of one of the representations in the first entry in the list above.

\subsection{Categories of D-branes}
\label{subsec: exceptional d-brane categories}

As a counterpart to the fusion ring bundle structure governing closed string states, we now turn to the D-brane categories. These provide a categorified structure—stacks of module categories over the fusion category—which capture how open string data and boundary conditions vary globally across the moduli space.

 The construction is very similar to the anomalous invertible symmetry in the compact boson example in Section~\ref{sect:genl:anom}. Recall that in that setup, the symmetry can be formulated as a fusion category Vec$_{\mathbb{Z}_2}^\omega$, over which D-branes correspond to objects in the module category. As elaborated in Section~\ref{sect:anom:modspace}, the anomaly $\omega$ of the $\mathbb{Z}_2$ symmetry implies there only exists a regular module category as Vec$_{\mathbb{Z}_2}^\omega$ itself, under which D-branes transform as two simple objects $M_+$ and $M_-$, understood as carrying nontrivial representations of $\mathbb{Z}_2$. The associativity data of the category is captured by the worldsheet boundary associator in Figure \ref{fig:twisted-z2-associator}.

Now our interested fusion categorical symmetry becomes those of the Ising and the tricritical Ising categories, for Spin$(7)$ and $G_2$, respectively. The D-branes as conformal boundaries of worldsheet SCFTs should again arranged as objects of associated module categories. We thus ask what are possible module categories for fusion categories under consideration. Mathematically, for a given fusion category $\mathcal{C}$, its module category $
\mathcal{M}$ is in one-to-one correspondence with its algebra object $A$ (see, e.g., \cite{Ostrik:2001xnt})
\begin{equation}
\mathcal{C}_A: \mathcal{M}~\text{as $A$-module category over}~\mathcal{C}.
\end{equation}
Physically, algebra object $A$ gives rise to a certain gauging of $\mathcal{C}$ symmetry in 2D QFTs, which for the invertible group symmetry $G$ is classified by subgroup $H\subset G$ and its discrete torsion $H^2(H, U(1))$.

Both Ising and tricritical Ising category have only a single algebra object $A=1$ (up to Morita equivalence\footnote{For Ising category, $A=1+\eta$ is also an algebra object, but will lead to the same module category as $A=1$. Similarly, for tricritical Ising with Fibonacci category as its subsector, $A=1+W$ is also an algebra object but equivalent to $A=1$.}), so the corresponding module category is just the regular one $\mathcal{C}_1\equiv \mathcal{C}$ as the fusion category itself\footnote{Beyond the regular module category, deriving $\mathcal{C}_A$ for nontrivial $A$ could be hard. We refer the reader to \cite{Yu:2025iqf} for one possible computational technique with concrete examples.}. Utilizing this result, we can obtain D-branes acted upon by the Ising and the tricritical Ising categorical symmetry as the following objects
\begin{equation}
\begin{split}
     \text{Spin}(7)&: M_1, ~M_\eta, ~M_\mathcal{D},\\
   G_2&: M_1, ~M_W, ~M_\eta, ~M_\mathcal{D}, ~M_{\eta W}, ~M_{\mathcal{D} W}.\\
\end{split}
\end{equation}

\begin{figure}[h]
    \centering
    \includegraphics[width=7cm]{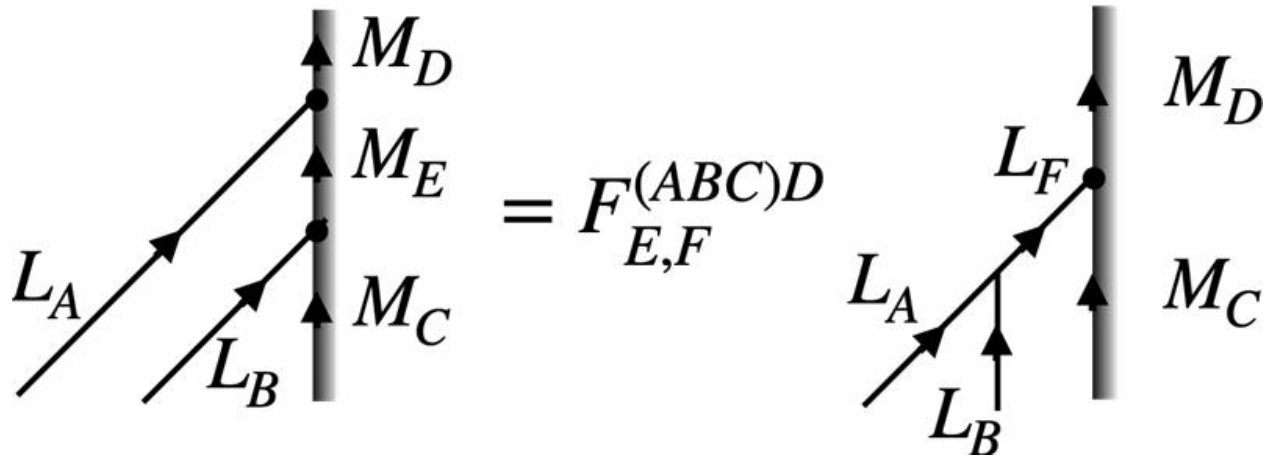}
    \caption{Boundary associators for the regular module category as F-symbols of the fusion category. Here the $A,B, \cdots$ labels simple objects $1,\eta, \mathcal{D}, W, \cdots$ in the Ising and the tricritical Ising categories.}
    \label{fig:Fsymbols}
\end{figure}

The actions non-invertible symmetries on D-branes can then be simply read from the fusion rules in (\ref{eq:fusion rule of Ising}) and (\ref{eq: fusion rule of tricritical Ising})
\begin{equation}
\begin{split}
    &\text{Spin}(7): \eta \otimes \begin{pmatrix}
        M_1\\
        M_\eta\\
        M_\mathcal{D}
    \end{pmatrix}=\begin{pmatrix}
        M_\eta\\
        M_1\\
        M_\mathcal{D}
    \end{pmatrix}, ~\mathcal{D}\otimes \begin{pmatrix}
        M_1\\
        M_\eta\\
        M_\mathcal{D}
    \end{pmatrix}=\begin{pmatrix}
        M_\mathcal{D}\\
        M_\mathcal{D}\\
        M_{1}+M_\eta
    \end{pmatrix}, ~\\
    &G_2: \eta \otimes \begin{pmatrix}
        M_1\\
        M_W\\
        M_\eta\\
        M_{\mathcal{D}}\\
        M_{\eta W}\\
        M_{\mathcal{D}W}
    \end{pmatrix}=\begin{pmatrix}
        M_\eta\\
        M_{\eta W}\\
        M_1\\
        M_{\mathcal{D}}\\
        M_{W}\\
        M_{\mathcal{D}W}
    \end{pmatrix}, ~\mathcal{D}\otimes \begin{pmatrix}
        M_1\\
        M_W\\
        M_\eta\\
        M_{\mathcal{D}}\\
        M_{\eta W}\\
        M_{\mathcal{D}W}
    \end{pmatrix}=\begin{pmatrix}
        M_\mathcal{D}\\
        M_{\mathcal{D} W}\\
        M_{\mathcal{D}}\\
        M_{1}+M_\eta\\
        M_{\mathcal{D} W}\\
        M_{W}+M_{\eta W}
    \end{pmatrix}, ~W \otimes \begin{pmatrix}
        M_1\\
        M_W\\
        M_\eta\\
        M_{\mathcal{D}}\\
        M_{\eta W}\\
        M_{\mathcal{D}W}
    \end{pmatrix}=\begin{pmatrix}
        M_W\\
        M_1+M_W\\
        M_{\eta W}\\
        M_{\mathcal{D}W}\\
        M_\eta+M_{\eta W}\\
        M_{\mathcal{D}}+M_{\mathcal{D}W}
    \end{pmatrix}.
\end{split}
\end{equation}
In order to have a stack over the moduli space, we also expect D-branes capturing the associativity data of the fusion category. This is indeed realized by the boundary associator in Figure \ref{fig:Fsymbols}, which in this case for regular module categories, are just F-symbols for the Ising and tricritical Ising categories. 
The concrete data of $F^{(ABC)D}_{E,F}$ can be expressed in terms of crossing kernels $\tilde{K}$ under the identity\footnote{We have used the fact that for Ising and tricritical Ising category all simple objects are their own orientation reversal, i.e., $A=\bar{A}$.}
\begin{equation}
    \tilde{K}^{A,D}_{B,C}(E,F) = F^{(ABC)D}_{E,F},
\end{equation}
whose explicit expression can be found in, e.g.,  \cite{Moore:1988qv, Chang:2018iay}. 

This explicit realization of module categories and their associator data provides concrete evidence for the general proposal of Section~\ref{sect:noninv}, where we conjectured that non-invertible symmetries organize D-brane categories into stacks fibered over moduli space.

\section{Conclusions}

In this work, we proposed that global symmetries, whether invertible, anomalous, or non-invertible, give rise to natural fiber structures over the moduli space of SCFTs. For closed string states, these structures form bundles of fusion rings or group rings; for D-brane categories, they form stacks of module categories or related categorical structures. This general framework extends the familiar Bagger-Witten line bundle story to theories without $U(1)_R$, particularly those with exceptional holonomy.

We illustrated this framework with the example of SCFTs arising from compactification on $G_2$ and Spin$(7)$ manifolds. In these cases, worldsheet non-invertible symmetries of (tricritical) Ising type organize the moduli space data via stacks of fusion categories and bundles of fusion rings, providing an analogue of the Bagger-Witten and Hodge line bundles in the absence of $U(1)_R$. These structures may reflect deeper categorified symmetry structures in target space, even though their precise spacetime interpretation remains elusive.

Our results suggest several future directions. To name a few:
\begin{itemize}
    \item Given the role of the (tricritical) Ising model in the worldsheet SCFT,
we also wonder if it makes an appearance in low-dimensional supergravity theories,
at leading order in string coupling, perhaps related to subtleties analogous to those which motivated the original Bagger-Witten paper \cite{Witten:1982hu}. 
\item It would be interesting to construct stacks we proposed in explicit examples, such as $G_2$ and Spin$(7)$ manifolds constructed by Joyce \cite{joyce1996compact1, joyce1996compact2, joyce1996compact3}.
\item A specific construction of Spin$(7)$ manifolds are via anti-holomorphic involution quotient of Calabi-Yau 4-folds \cite{joyce1999new}, due to the fact every Calabi-Yau 4-fold can be regarded as a Spin$(7)$ manifold. From the worldsheet perspective, the extended $\mathcal{N}=1$ algebra  of Spin$(7)$ can be embedded into the extended $\mathcal{N}=2$ algebra of Calabi-Yau, and D-branes calibrated by Cayley 4-forms are realized as $\frac{1}{4}$-BPS boundaries of the Calabi-Yau SCFTs \cite{DYY}. It would be interesting to see how the Ising categorical symmetry of the Spin$(7)$ can be embedded according in Calabi-Yau 4-fold SCFTs, and how this embedding interplay with the Bagger-Witten line bundle for the Calabi-Yau.
\end{itemize}

\acknowledgments
We would like to thank L.~Anderson, R.~Bryant, 
D.~BenZvi, M.~Bullimore, R.~Donagi, J.~Gray, H.~T.~Lam, J.~McNamara, I.~Melnikov, T.~Pantev, Y.~Tachikawa, E.~Torres, and H.~Y.~Zhang
for useful discussions. We thank Y.~Tachikawa for pointing out eariler work on 2-group structure in string theory. X.Y. also thanks H.~T.~Lam for raising the question of the spacetime interpretation of the $U(1)_m\times U(1)_w$ anomaly, and J.~McNamara for pointing out relevant references and fruitful discussion of it. 
E.S.~and X.Y.~were partially supported by NSF grant PHY-2310588.

\appendix

\section{Technical remarks} \label{app:fusion-2gp}

\subsection{Preliminaries on stacks and 2-vector bundles}
\label{app:stack-preliminaries}
A stack is an analogue of a bundle or sheaf, but in which the fibers or sections are categories rather than sets, rings, or modules.  See for example \cite{BBK06} for the case that the category $\cal C$ is monoidal but not necessarily linear. The more classical case of fusion categories coming from Hopf algebras is more standard in the context of noncommutative geometry. Some references on this include \cite{dadp,skoda,Brzezinski:1992hda,brz,ugalde,durdevic,aflw}.

One can also consider 2-vector bundles associated with these higher principal bundles, in analogy to working with associated complex line bundles instead of principal $U(1)$ bundles. In this case, one considers taking as a fiber a complex linear category $\cal M$ with a $\cal C$-module structure. As $\cal M$ is an object in the symmetric monoidal 2-category of abelian complex linear categories \cite{Fra13}, it is a 2-vector space (see e.g. \cite[Definition 7.1]{FHLT09}). In particular, one can take ${\cal M}={\cal C}$, the underlying linear category of $\cal C$ obtained by forgetting its fusion product. This is indeed the case of interest for the CFT's appearing in this paper, as the primary operators are in one-to-one correspondence with the Verlinde lines.
A definition of bundles with 2-vector spaces as fibers, namely 2-vector \textit{bundles}, is provided in \cite{KLW21,KLW22}. It can be stated in terms of Čech data of algebra bundles. 

As we do not attempt to compute examples in this paper, we leave such technical considerations for future work.

\subsection{Fusion categories and 2-groups}   

In the text we often utilize the fusion category Vec$(G,\alpha)$ for finite groups $G$,
to describe a (zero-form) symmetry $G$ in a two-dimensional theory with a
't Hooft anomaly $\alpha \in H^3(G,U(1))$.  Now, it is also sometimes said 
(see e.g.~\cite[section 5]{Sharpe:2015mja}, \cite{Cordova:2018cvg,Benini:2018reh,Cordova:2020tij,DelZotto:2020sop,Iqbal:2020lrt})
that such symmetries should be described in terms of 2-group extensions of the form
\begin{equation}
    1 \: \longrightarrow \: BU(1) \: \longrightarrow \: \tilde{G} \: \longrightarrow \:
    G \: \longrightarrow \: 1.
\end{equation}
Although the languages seem different, these are two descriptions of what should ultimately be the same symmetry.  In this appendix, we describe the relationship.

    To see this, we can first describe the 2-groups as monoidal categories. The 2-group $BU(1)$ is described as a monoidal category $\underline{BU(1)}$ with a single object $\bullet\in \text{ob}(\underline{BU(1)})$, the unique monoidal product, and morphisms $\underline{BU(1)}(\bullet,\bullet)=U(1)$. The 2-group extension $\tilde{G}$ can be described as the monoidal category $\underline{\tilde{G}}$ with objects $\text{ob}(\underline{\tilde{G}})=G$, morphisms $\underline{\tilde{G}}(g,h)=\delta_{g,h}\, U(1)$ (where the delta function for $g\neq h$ means the hom-set is the empty set), and the monoidal product is given by the group law of $G$ taken to be associative up to the extension 3-cocycle $\alpha \in Z^3(G,U(1))$. In other words, the monoidal category $\underline{\tilde{G}}$ comes equipped with nontrivial associators
    \begin{equation}
        \alpha_{g,h,k}: (g\otimes h)\otimes k \xrightarrow{\cong} g\otimes (h\otimes k),
    \end{equation}
    satisfying the usual associativity condition as a consequence of the 3-cocycle condition that $\alpha$ satisfies.
    Finally, the group $G$ is described as the monoidal category $\underline{G}$ with objects labeled by $G$, and only identity morphisms $\underline{G}(g,h)=\delta_{g,h} \,1$ (where here $1=\{e\}$ is the trivial group). These monoidal categories come equipped with monoidal functors
    \begin{equation}
        \underline{BU(1)} \xrightarrow{\imath} \underline{\tilde{G}}\xrightarrow{\pi} \underline{G},
    \end{equation}
    where $\imath(\bullet)=1\in \text{ob}(\underline{\tilde{G}})$, $\imath(z)=z\in \underline{\tilde{G}}(1,1)=U(1)$ for $z\in \underline{BU(1)}(\bullet,\bullet)$, $\pi(g)=g\in \text{ob}(\underline{G})$, and $\pi(z)=e\in \underline{G}(g,g)=1$. The image of the composition $\pi\circ\imath: \underline{BU(1)}\to \underline{G}$ is trivial both on the objects and morphisms, as expected.

    These monoidal categories admit embeddings into the tensor categories of graded vector spaces
    \begin{equation}
\begin{tikzcd}
\underline{BU(1)} \arrow[rr, "\imath"] \arrow[dd, hook] &  & \underline{\tilde{G}} \arrow[rr, "\pi"] \arrow[dd, hook] &  & \underline{G} \arrow[dd, hook] \\
                                                        &  &                                                          &  &                                \\
\text{Vec} \arrow[rr, "\imath_{\rm tensor}"', hook]     &  & {\text{Vec}(G,\alpha)} \arrow[rr, "\pi'"']               &  & \text{Vec}(G)                 
\end{tikzcd},
    \end{equation}
where, for example, an object $g\in\text{ob}(\underline{\tilde{G}})$ is sent to the homogeneous $G$-graded vector space $U_g$ of degree $g\in G$, and the morphisms embed via the inclusion $U(1)\hookrightarrow\mathbb{C}^{\times}$. This sequence is understood as the linear extension of the 2-group exact sequence. While the linear extension $\pi'$ of the functor $\pi$ is in general not a tensor functor (as it is rarely faithful), the linear extension $\imath_{\rm tensor}$ of $\imath$, as the notation suggests, is a tensor embedding of tensor categories. This is the information one obtains via the extension of fusion categories using homotopy-theoretic techniques developed in \cite{Etingof:2009yvg}. 

In summary, $\text{Vec}(G,\alpha)$ is simply the linear extension of $\tilde{G}$, as it is constructed from the same information, namely a finite group $G$ and an extension class $[\alpha]\in H^3(G,U(1))$. Intuitively, this can be understood as the once-categorified construction of the complex group algebra $\mathbb{C}[G]$ of a finite group $G$. In the case of group algebras, to the group $G$ we assign a vector space $\mathbb{C}[G]$ such that each group element $g\in G$ has a corresponding basis element $v_g\in \mathbb{C}[G]$. These basis elements form a complete basis. Furthermore, the vector space inherits the group multiplication from $G$. In our case of interest, to the 2-group $\tilde{G}$ we assign an abelian \textit{category} $\mathcal{C}$, such that to each object $g\in G$ there corresponds a different irreducible object $U_g\in\text{ob}(\mathcal{C})$ (the analogue of a basis element), so that every object in $\cal C$ is the sum of these irreducible elements (the analog of a complete basis). But now there is an additional layer of information, the $U(1)$ phases that are morphisms in $\tilde{G}$. Hence, to each of these we assign a morphism in $\cal C$. The abelian category $\mathcal{C}$ inherits the 2-group multiplication from $\tilde{G}$, which is that of $\text{Vec}(G,\alpha)$.

Finally, note that we are speaking of \textit{the} linear extension in the sense that it is the $\mathbb{C}$-linear context of lowest category theoretic degree that faithfully incorporates the (higher) group products. However, there are more general linear extensions, which become relevant when talking about finite non-invertible symmetries of higher-dimensional theories. The clearest example is $n\text{Vec}(G)$ the fusion $n$-category of $G$-graded $n$-vector spaces. This is definitely a linear extension of $G$, albeit of a potentially high category theoretic degree, which is thought to encode a $0$-form $G$-symmetry in a $(n+1)$-dimensional theory. 

\subsection{Gaugeability of fusion category symmetries in 2D}
\label{app:gaugeability}
In two-dimensional quantum field theories, not every fusion category symmetry can be gauged. A necessary and sufficient condition for the gauging of a fusion category $\mathcal{C}$ is the existence of a \emph{special symmetric Frobenius algebra} object in $\mathcal{C}$, as discussed in, e.g., \cite{Fuchs:2002cm,Choi:2023vgk, Perez-Lona:2023djo,Perez-Lona:2024sds,Diatlyk:2023fwf}. This condition is related to, but distinct from, other categorical notions such as fiber functors or regular objects.

The latter are objects of the form
\begin{equation} \label{eq:defn-reg2}
    R \: = \: \bigoplus_L (\dim L) \, L,
\end{equation}
where the sum is over all simple objects $L$ in the fusion category, and the coefficients (the quantum dimensions of the simple objects)
are required to be integers.  
\begin{itemize}
\item To gauge the action of a fusion category in two dimensions, one must specify a special symmetric Frobenius algebra object.
    \item If the fusion category admits a fiber functor, then the quantum dimensions are automatically integers, and in addition it admits a special symmetric Frobenius algebra object.  As a result, the fusion category can be gauged.
    \item Regular objects can still exist in other fusion categories for which simple objects have integer quantum dimension, but there is no fiber functor, and no symmetric special Frobenius algebra structure.  An example is Vec$(G,\alpha)$ for $G$ a finite group and $\alpha$ a nontrivial element of $H^3(G,U(1))$.  The example Vec$(G,\alpha)$ is not gaugeable for nontrivial $\alpha$, corresponding mathematically to the fact that there is no symmetric special Frobenius algebra structure on the regular object, and physically to the fact that $\alpha$ is an anomaly which obstructs gauging. Further examples of this kind are described in \cite{Perez-Lona:2025ncg} as representation categories of quasi Hopf algebras.
    \item There are also further fusion categories for which at least one simple object does not have integer quantum dimension, hence no regular object exists (as one cannot write a linear combination of
the form~(\ref{eq:defn-reg2}) with integer coefficients).  Examples include Ising and Fibonacci.
Some special cases (with objects of non-integer quantum dimension) still admit symmetric special Frobenius algebra objects, and so one can gauge even if one does not have a fiber functor.    
\end{itemize}
In any event, neither having a fiber functor nor having a regular object are necessary for gaugeability.

\bibliographystyle{utphys}
\bibliography{draft}
\end{document}